\documentclass[12pt]{article}

\usepackage[active]{srcltx}
\usepackage{amsfonts,amssymb,amsmath}
\usepackage{paralist}
\usepackage{graphicx}
\usepackage{graphics} 
\usepackage{epsfig} 

\usepackage{hyperref} 

\usepackage{floatflt}
\usepackage{psfrag}
\usepackage{color}

\newtheorem{theorem}{Theorem}[section]

\newtheorem{proposition}{Proposition}

\newtheorem{definition}[theorem]{Definition}
\newtheorem{remark}{Remark}

\title{Two-way multi-lane traffic model for pedestrians in corridors}

\author{C\'{e}cile Appert-Rolland, Pierre Degond and S\'{e}bastien Motsch}

\begin{document}
\maketitle

\centerline{\scshape C\'{e}cile Appert-Rolland}
\medskip
{\footnotesize
  \centerline{1-University Paris-Sud; Laboratory of Theoretical Physics}
  \centerline{Batiment 210, F-91405 ORSAY Cedex, France.}
} 
{\footnotesize
  \centerline{2-CNRS; LPT; UMR 8627}
  \centerline{Batiment 210, F-91405 ORSAY Cedex, France.}
}
{\footnotesize
  \centerline{email: Cecile.Appert-Rolland@th.u-psud.fr}
}

\medskip

\centerline{\scshape Pierre Degond}
\medskip
{\footnotesize
  \centerline{3-Universit\'e de Toulouse; UPS, INSA, UT1, UTM}
  \centerline{Institut de Math\'ematiques de Toulouse}
  \centerline{F-31062 Toulouse, France.}
} %
{\footnotesize
  \centerline{4-CNRS; Institut de Math\'ematiques de Toulouse UMR 5219}
  \centerline{F-31062 Toulouse, France.}
} %
{\footnotesize
  \centerline{email: pierre.degond@math.univ-toulouse.fr}
}

\medskip

\centerline{\scshape S\'{e}bastien Motsch}
\medskip
{\footnotesize
  \centerline{5-Department of Mathematics University of Maryland}
  \centerline{College Park, MD 20742-4015, USA.}
} %
{\footnotesize
  \centerline{email: smotsch@cscamm.umd.edu}
}



{\bf Acknowledgements:} This work has been supported by the french 'Agence Nationale pour la Recherche
  (ANR)' in the frame of the contract 'Pedigree' (contract number
  ANR-08-SYSC-015-01). The work of S. Motsch is partially supported by NSF grants
  DMS07-07949, DMS10-08397 and FRG07-57227.
\bigskip

\begin{abstract}
  We extend the Aw-Rascle macroscopic model of car traffic into a two-way multi-lane model
  of pedestrian traffic. Within this model, we propose a technique for the handling of the
  congestion constraint, i.e. the fact that the pedestrian density cannot exceed a maximal
  density corresponding to contact between pedestrians. In a first step, we propose a
  singularly perturbed pressure relation which models the fact that the pedestrian
  velocity is considerably reduced, if not blocked, at congestion. In a second step, we
  carry over the singular limit into the model and show that abrupt transitions between
  compressible flow (in the uncongested regions) to incompressible flow (in congested
  regions) occur. We also investigate the hyperbolicity of the two-way models and show
  that they can lose their hyperbolicity in some cases. We study a diffusive correction of
  these models and discuss the characteristic time and length scales of the instability.
\end{abstract}

\section{Introduction}

Crowd modeling and simulation is a challenging problem which has a broad range of applications from public safety to entertainment industries through architectural and urban design, transportation management, etc. Common and crucial needs for these applications are the evaluation and improvement (both quantitatively and qualitatively) of existing models, the derivation of new experimentally-based models and the construction of hierarchical links between these models at the various scales. 

The goal of this paper is to propose a phenomenological macroscopic model for pedestrian movement in a corridor. A macroscopic model describes the state of the crowd through locally averaged quantities such as the pedestrian number density, mean velocity, etc. Macroscopic models are opposed to Individual-Based Models (IBM's) which follow the location and state of each agent over time. Macroscopic models provide a description of the system at scales which are large compared to the individuals scale. Although they do not provide the details of the individuals scale, they are computationally more efficient. In particular, their computational cost does not depend on the number of agents, but only on the refinement level of the spatio-temporal discretization. In addition, by comparisons with the experimental data, they give access to large-scale information about the system. This information can provide a preliminary gross analysis of the data, which in turn can be used for building up more refined IBM's. This procedure requires that the link between the microscopic IBM and the macroscopic model has been previously established. Therefore, macroscopic models which can be rigorously derived from IBM's are crucial.

The present work focuses on a one-dimensional model of pedestrian traffic in corridors. This setting has several advantages: 
\begin{enumerate}
\item It makes the problem essentially one-dimensional and is a preliminary step for the development of more complex multi-dimensional problems. The present work will consider that pedestrian traffic occurs on discrete lanes. This approximation can be viewed as a kind of discretization of the actual two-dimensional dynamics. It prepares the terrain for the development and investigation of truly two-dimensional models. 
\item We can build up on previous experience in the field of traffic flow models. Our approach relies on the Aw-Rascle model of traffic flow \cite{AR}, which has been proven an excellent model for traffic flow engineering \cite{Zhang}. In the present work, this approach will be extended to two-way multi-lane traffic flow of pedestrians.
\item It is easier to collect well-controlled experimental data in corridors than in open space (see for instance \cite{Ped1}). 
\item The relation of the macroscopic model to a corresponding microscopic IBM is more easily established in the one-dimensional setting. In \cite{AKMR}, it has been proven that the Aw-Rascle model can be derived from a microscopic Follow-the-Leader model of car traffic. The proof uses a Lagrangian formulation of the Aw-Rascle model. The correspondence between the Lagrangian formulation and the IBM cannot be carried over to the two dimensional case, because of the very special structure of the Lagrangian model in one-dimension. 
\end{enumerate}

The most widely used models of pedestrian traffic are IBM's. Several families of models
have been developed.  Rule based models \cite{Reynolds99} have been used in particular for
the development of games and virtual reality, with several possible levels of
description. But their aim is more to have a realistic appearance rather than really
reproducing a realistic behavior.  More robust models are needed for example to test and
improve the geometry of various types of buildings.  Physicists have proposed some models
inspired from the fluid simulation methods.  In the so-called 'social force' model
\cite{Helbing_1991,Helbing_Molnar_1995,Helbing_Molnar_1997}, the equations of motion for
each pedestrian have the form of Newton's law where the force is the sum of several terms
each representing the 'social force' under consideration. It obviously relies on the
analogy existing between the displacement of pedestrians and the motion of particles in a
gas. It describes quite well dense crowds, but not the individual trajectories of a few
interacting pedestrians.  Other approaches have been developed in the framework of
cellular automata \cite{Burstedde_2001,Guo_2008,Nishinari_2004}. In these models, the
non-local interactions between pedestrians are made local through the mediation of a
virtual floor field. These models also are meant to describe the motion of crowds, not of
individuals.  Besides, a systematic study of the isotropy of cellular automata models is
still lacking.  More recently, some geometrical models have been developed. Pedestrians
try to predict each others' trajectories, and to avoid collisions
\cite{Guy09, Paris_p_d07,Vandenberg_o08}.  The knowledge of other pedestrians' trajectories
depends on the perception that the pedestrian under consideration has, which may vary with
time.  \cite{Pettre09} takes into account the fact that this knowledge is acquired
progressively.  Another type of perception based on the visual field is proposed in
\cite{Ondrej10}.  These models describe well the individual trajectories of a few
interacting pedestrians, but it is not obvious yet whether they can handle crowds.

By contrast to microscopic IBM's, macroscopic crowd models are based on the analogy of
crowd flow with fluid dynamics. A first approach has been proposed in
\cite{Henderson_1974}. In \cite{Helbing_1992}, a fluid model is derived from a gas-kinetic
model through a moment approach and phenomenological closures. Recently, a similar
approach has been proposed in \cite{Al-nasur_2006}. In \cite{Hoogendoorn_2003,
  Hughes_2002, Hughes_2003}, a continuum model is derived through optimal control theory
and differential games. It leads to a continuity equation coupled with a potential field
which describes the velocity of the pedestrians.  Other phenomenological models based on
the analogy with the Lighthill-Whitham-Richards model of car traffic have been proposed by
\cite{Bellomo_d08, Chalons_2007, Colombo_2005}.  In \cite{Piccoli_2009, Piccoli_2010},
instead of considering a continuous time evolution described by PDE's, the evolution of
measures is performed on a discrete time scale. In the present paper, we shall consider a
continuous time description. Macroscopic models provide a description of the system at large spatial scales. They can be heuristically justified for a long corridor stretch like a subway corridor, when the spatial inhomogeneities are weak (such as low variations of the density or velocity in the direction of the corridor). Of course, they cannot be used when the spatial inhomogeneities are at the same scale as for instance the mean-interpedestrian distance in the longitudinal direction. In the case of narrow corridors, this mean-interpedestrian distance is larger because there are less pedestrians in a cross-section, and the condition of weak spatial inhomogeneities is more stringent. From a rigorous standpoint, the derivation of macroscopic models from Individual-Based models requires that the number of agents be large, which is obviously questionable in most situations in pedestrians and highway traffic. Still there is a large literature devoted to macroscopic models which seem to provide adequate models for large scale dynamics.

We will be specifically interested in two-way multi-lane traffic flow models with a
particular emphasis on the handling of congestions. These points have been previously
addressed in \cite{Weng_2007} for pedestrian counter-flows, \cite{Shvetsov_1999} for
multi-lane traffic and \cite{Maury_Roudneff_2010, Maury_Venel_2008} for the treatment of
congestions. However, to the best of our knowledge, none of these different features have
been included in the same model at the same time. The most difficult point is the
treatment of congestions. In the recent approach \cite{Maury_Roudneff_2010,
  Maury_Venel_2008} the congestion constraint (i.e. the limitation of the density by a
maximal density corresponding to contact between pedestrians) is enforced by means of
convex optimization tools (for IBM's) or techniques borrowed from optimal transportation
such as Wasserstein metrics (for continuum models). However, these abstract methods do not
leave much space for parameter fitting to data and cannot distinguish between the behavior
of pedestrians and say, sheep. Our technique relies on the explicit derivation of the
dynamics of congestions, in the spirit of earlier work for traffic
\cite{BDDR,BDLMRR,Deg_Del}. This procedure was initiated in the seminal work
\cite{Bou_Bre_Cor_Rip}.

The outline of the paper is as follows. We first present the modeling approach for a one-way
one-lane Aw-Rascle model (1W-AR) of pedestrian flow in corridors in section
\ref{sec_1lane_1way}. We then successively extend this model into a two-way one-lane Aw-Rascle model (2W-AR) in section
\ref{1lane} and to a two-way multi-lane Aw-Rascle model (ML-AR) in section \ref{mlane}. In each section, we
present the corresponding Aw-Rascle model, together with a simplified version of it supposing that the pedestrian desired velocity is constant and uniform. We refer these simplified models as ''Constant desired velocity Aw-Rascle'' (CAR) models. Therefore, we successively have the 1W-CAR, 2W-CAR and ML-CAR models as Constant desired velocity versions of respectively the 1W-AR, 2W-AR and ML-AR models. The 1W-CAR model can be recast in the form of the celebrated Lighthill-Whitham-Richards (LWR) model of traffic. 

Finally, for each of these models, we propose a specific treatment of congestion regions. This treatment consists in introducing a singular pressure in the AR model which tends to infinity as the density approaches the congestion density (i.e. the density at which the agents are in contact to each other). This singularly perturbed pressure relation provides a significant reduction of the flow when the density reaches this maximal density. A small parameter $\varepsilon$ controls the thickness of the transition region. In the limit $\varepsilon \to 0$, two phases appear: an uncongested phase where the flow is compressible and a congested phases where the flow is incompressible. The transition between these two phases is abrupt, by contrast to the case where $\varepsilon$ stay finite, where this transition is smooth. The location of the transition interface is not given a priori and is part of the unknowns of the limit problem. 

Table \ref{table_1} below provides a summary of the various proposed models and their relations.

\begin{table}[ht]
\begin{center}
\begin{tabular}{|l||c|c|c||}
\hline
   & Basic model  & Congestion model    & Congestion model    \\
             &    & with smooth transition:  & with abrupt transition:  \\
             &    & finite $\varepsilon$  & $\varepsilon \to 0$  \\
\hline
\hline
1W-AR             & Section               & Section              & Section               \\
1-way          & \ref{subsec_AR_model} & \ref{subsubsec_AR_density_constraint_smooth} &  \ref{subsubsec_AR_density_constraint_incompressibility}  \\
1-lane & & & \\
\hline
1W-CAR             & Section              & Section              & Section        \\
1-way            & \ref{subsec_toy} & \ref{subsubsec_TAR_density_constraint} & \ref{subsubsec_TAR_density_constraint} \\
1-lane & & item (i) & item (ii) \\
\hline
\hline
2W-AR             & Section               & Section              & Section               \\
2-way          & \ref{subsec_2AR_model} & \ref{subsubsec_2WAR_density_constraint_smooth} &  \ref{subsubsec_2WAR_density_constraint_incompressibility}  \\
1-lane & & & \\
\hline
2W-CAR             & Section              & Section              & Section        \\
2-way            & \ref{subsec_2TAR_model} & \ref{subsubsec_T2AR_density_constraint} & \ref{subsubsec_T2AR_density_constraint} \\
1-lane & & item (i) & item (ii) \\
\hline
\hline
ML-AR             & Section               & Section              & Section               \\
2-way          & \ref{subsec_mlane_principles} & \ref{subsubsec_kWAR_density_constraint_smooth} &  \ref{subsubsec_kWAR_density_constraint_incompressibility}  \\
multi-lane & & & \\
\hline
ML-CAR             & Section              & Omitted             & Omitted        \\
2-way            & \ref{subsec_mlane_toy} &  &  \\
multi-lane & &  &  \\
\hline
\hline
\end{tabular}
\vspace{0.2cm} \\
\caption{\small Table of the various models, with some of their characteristics and the sections in which they are introduced. The meaning of the acronyms is as follows: AR='Aw-Rascle model', CAR='Aw-Rascle model with Constant Desired Velocity'. The left column (basic model) refers the general formulation of the model and the middle and right columns, the modified models taking into account the congestion phenomena. The middle column corresponds to a smooth transition from uncongested state to congestion while the right column corresponds to an abrupt phase transition. }
\label{table_1} 
\end{center}
\end{table}

One interesting characteristics of two-way models as compared to one-way models is that
they may lose their hyperbolicity in situations close to the congestion regime. Although,
this loss of hyperbolicity can be seen as detrimental to the model, the
resulting instability may explain the appearance of crowd turbulence at high
densities. We note that a loss of (strict) hyperbolicity has already been found 
in a multi-velocity one-way model \cite{benzoni}. 
In order to gain insight into this instability, in section \ref{sec_math}, we
analyze the diffusive perturbation of the two-way Aw-Rascle model with constant desired
velocity, and exhibit the typical time scale and growth rate of the so-generated
structures. These observables can be used to assess the model and calibrate it against
empirical data. In order to illustrate these considerations, we show numerical simulations
that confirm the appearance of these large-scale structures which consist of two
counter-diffusing crowds. In these simulations, which are presented for illustrative purposes only, in order to explore what kind of structures the lack of hyperbolicity of the model leads to, we assume a smooth pressure-density relation. Thanks to this assumption, we omit to treat the congestion constraint, which is a difficult stiff problem, for which special methods have to be designed (see e.g. \cite{DT, DHN} for the case of Euler system of gas
dynamics and \cite{BDDR, Deg_Del} for the AR model).

\section{One-way one-lane traffic model}
\label{sec_1lane_1way}

\subsection{An Aw-Rascle model for one-lane one-way pedestrian traffic}
\label{subsec_AR_model}

In this section, we construct a one-lane one-way continuum model of pedestrian traffic in corridors. In this model, we will pay a particular attention to the occurrence of congestions. We encode the congestion effect into a constraint of maximal total density. This work is inspired by similar approaches for vehicular traffic, which have been developed in \cite{BDDR,BDLMRR,Deg_Del}. 

For that purpose, the building block is a one-lane, one-way Aw-Rascle (1W-AR) model which has been proposed for vehicular traffic flow \cite{AR}. This model belongs to the class of second-order models in the sense that it considers that both the density and the velocity are dynamical variables which are subject to time-differential equations. By contrast, first order models use the density as the only dynamical variable and prescribe the density flux as a local function of the density. The Aw-Rascle model with constant desired velocity considered in section \ref{subsec_toy} is an example of a first order model.

\begin{definition} {\bf (1W-AR model)} Let $\rho(x,t) \in {\mathbb R}$ the density of
  pedestrians on the lane, $u\in {\mathbb R}_+$ their velocity, $w(x,t) \in {\mathbb R}_+$
  the desired velocity of the pedestrian in the absence of obstacles and $p(\rho)$ the
  velocity offset between the desired and actual velocities of the pedestrian. The 1W-AR model is written:
  \begin{eqnarray}
    & & \hspace{-1cm} \partial_t \rho + \partial_x (\rho u) = 0 , \label{AR_n} \\
    & & \hspace{-1cm} \partial_t (\rho w)  + \partial_x (\rho w u) = 0 , \label{AR_u} \\
    & & \hspace{-1cm} w= u + p(\rho). \label{AR_w}
  \end{eqnarray}
\end{definition}
In this model, the offset $p(\rho)$ is an increasing function of the pedestrian density. By analogy with fluid mechanics, this offset will be often referred to as the pressure, but its physical dimension is that of a velocity.

Using the mass conservation equation, we can see that the desired velocity is a Lagrangian quantity (i.e. is preserved by the flow), in the sense that: 
\begin{eqnarray}
  & & \hspace{-1cm} \partial_t w  + u \partial_x w = 0  . \label{AR_w_lag} 
\end{eqnarray}
It is natural, since the desired velocity is a quantity which is attached to the particles and should move together with the particles at the flow velocity.

This model has been studied in great detail in \cite{AR} and proven to derive from a follow-the-leader model of car traffic in \cite{AKMR}. Of particular interest is the fact that this model is hyperbolic, with two Riemann invariants. The first one is obviously the desired velocity $w$ as (\ref{AR_w_lag}) testifies. The second one is less obvious but is nothing but the actual flow velocity $u$. Indeed, from (\ref{AR_w_lag}) and using (\ref{AR_n}), we get: 
\begin{eqnarray*}
  \partial_t u  + u \partial_x u &=& - (\partial_t p  + u \partial_x p)   \\
  &=& - p'(\rho) (\partial_t \rho  + u \partial_x \rho)  \\
  &=&  p'(\rho) \rho \partial_x u , 
\end{eqnarray*}
and therefore 
\begin{eqnarray}
  & & \hspace{-1cm} \partial_t u  + (u - p'(\rho) \rho) \partial_x u = 0  . \label{AR_u_lag}
\end{eqnarray}
Therefore, information about the fluid velocity propagates with a velocity 
\begin{eqnarray}
  & & \hspace{-1cm} c_u = u - p'(\rho) \rho. \label{AR_u_vel}
\end{eqnarray}
In the reference frame of the fluid, this gives raise to waves moving upstream the flow with a speed equal to $- p'(\rho) \rho$. 

\begin{remark}
  We can also consider the evolution of $\rho u$ instead of that of $u$. We obtain from (\ref{AR_w_lag}) and using (\ref{AR_n}): 
  \begin{eqnarray}
    \partial_t (\rho u)  + \partial_x (\rho u u ) &=& - (\partial_t (\rho p)   + \partial_x (\rho p u) )  \nonumber  \\
    &=& - \rho (\partial_t p  + u \partial_x  p) \nonumber  \\
    &=&  - \rho \frac{dp}{dt}  , \label{AR_rho_u}
  \end{eqnarray}
  where we have introduced the material derivative $d/dt = \partial_t + u \partial_x$. This form is motivated by the observation \cite{AR} that drivers do not react to local gradients of the vehicle density but rather to their material derivative in the frame of the driver. This modification to standard gas dynamics like models of traffic was crucial in obtaining a cure to the various deficiencies of second order models as observed by Daganzo \cite{Dag}. Eq.  (\ref{AR_rho_u}) can also be put in the form
  \begin{eqnarray}
    \partial_t (\rho u)  + \partial_x (\rho u w ) &=& - \partial_t (\rho p)     \nonumber  \\
    &=& - \pi'(\rho) \partial_t \rho \nonumber  \\
    &=&  \pi'(\rho) \partial_x (\rho u) , \label{AR_rho_u_2}
  \end{eqnarray}
  with 
  \begin{eqnarray}
    & & \hspace{-1cm} \pi(\rho) = \rho p(\rho), \quad \pi'(\rho) = \rho p'(\rho) + p(\rho) . \label{AR_pi} 
  \end{eqnarray}
\end{remark}

We will consider the 1W-AR model as a building block for the pedestrian model. In order to make the connection with a microscopic view of pedestrian flow, we consider a subcase of this model in the section below.

\subsection{Constant desired velocity}
\label{subsec_toy}

This one-way Constant Desired Velocity Aw-Rascle (1W-CAR) model assumes that the pedestrians can have only two velocities: either a fixed uniform velocity $V$ which is the same for all pedestrians and does not vary with time ; or zero, indicating that they are immobile. In other words, if because of the high density of obstacles in front, the pedestrians cannot proceed further with the velocity $V$, they have to stop. 

In this case, 
\begin{eqnarray*}
  & & \hspace{-1cm} w = V 
\end{eqnarray*}
is a fixed value and therefore, the actual flow velocity 
\begin{eqnarray}
  & & \hspace{-1cm} u = V  - p(\rho)\label{ARPed_u}
\end{eqnarray}
is a local function of $\rho$. 
This leads to a first-order model where the flux velocity is given as a local prescription
of the density.

\begin{definition} {\bf (1W-CAR model)} Let $\rho(x,t) \in {\mathbb R}$ the density of
  pedestrians on the lane, $V \in {\mathbb R}_+$ the (constant) desired velocity of
  pedestrians and $p(\rho)$ the pressure. The 1W-CAR model is written:
  \begin{eqnarray}
    & & \hspace{-1cm} \partial_t \rho + \partial_x (\rho (V - p(\rho)) ) = 0. \label{ARPed_n}
  \end{eqnarray}
\end{definition}

\begin{figure}
  \centering
  \includegraphics[scale=.7]{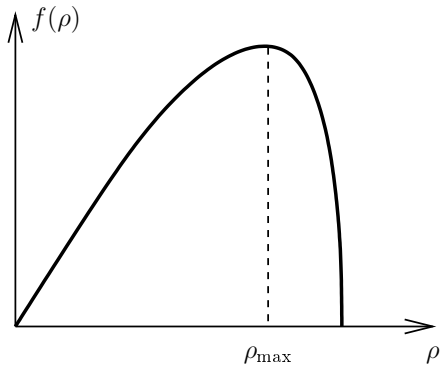}
  \caption{Density flux $f(\rho) = \rho (V - p(\rho))$ as a function of $\rho$ in the
    1W-CAR model.}
  \label{Fig_LWR}
  \label{fig:nom}
\end{figure}

We denote by $f(\rho) = \rho (V - p(\rho))$ the mass flux. The quantity $p(\rho)$ being an increasing function of $\rho$, $f(\rho)$ has a concave shape (and is actually concave if $\rho p(\rho)$ is convex), which is consistent with classical first-order traffic models such as the Lighthill-Whitham-Richards (LWR) model \cite{LW}. Figure \ref{Fig_LWR} provides a graphical view of $f(\rho)$. It is interesting to note that the original 1W-AR model can be viewed as a LWR model with a driver-dependent flux function $f(\rho,w) = \rho (w-p)$ where $w$ is the driver dependent parameter, and consequently moves with the flow speed. It follows that the LWR is a useful lab to test concepts ultimately applying to the 1W-AR model. However, some of the features of the LWR model are too simple (such as the conservation of the maxima and minima of $\rho$) and a realistic description of the dynamics requires more complex models such as the 1W-AR model.

It is also instructive to write the 1W-CAR model as a second order model, like the 1W-AR model. Indeed, using (\ref{AR_rho_u_2}) and (\ref{ARPed_u}), we can write (\ref{ARPed_n}) as: 
\begin{eqnarray}
  & & \hspace{-1cm} \partial_t \rho + \partial_x (\rho u) = 0 , \label{RARPed_n} \\
  & & \hspace{-1cm} \partial_t (\rho u)  + \partial_x (\rho u V ) = \pi'(\rho) \partial_x (\rho u) . \label{RARPed_u}
\end{eqnarray}
Conversely, if $\rho$ and $u$ are solutions of this model, using the fact that $V$ is a constant together with eq. (\ref{RARPed_n}) to modify the second term of (\ref{RARPed_u}), and using the r.h.s. of eq. (\ref{AR_rho_u_2}) to modify the r.h.s of (\ref{RARPed_u}), we find, : 
\begin{eqnarray*}
  & & \hspace{-1cm} \partial_t (\rho (p + u - V) ) = 0. 
\end{eqnarray*}
Therefore, if (\ref{ARPed_u}) is satisfied initially, it is satisfied at all times and we recover (\ref{ARPed_n}). 

\begin{remark}
  The 1W-CAR model in the form
  (\ref{RARPed_n}), (\ref{RARPed_u}) has an interesting interpretation in terms of
  microscopic dynamics, when the pedestrians have two velocity states, the moving one with
  velocity $V$ and the steady one, with velocity $0$. Indeed, denoting by $g(x,t)$ the
  density of moving pedestrians and by $s(x,t)$ that of steady pedestrians, we have
  \begin{eqnarray*}
    & & \hspace{-1cm} \rho = g + s  . 
  \end{eqnarray*}
  Because the moving pedestrians move with velocity $V$, we can write the pedestrian flux $\rho u$ as
  \begin{eqnarray}
    & & \hspace{-1cm} \rho u = V g  . \label{RARPed_rhou=Vg}
  \end{eqnarray}
  Since by (\ref{ARPed_n}), $\rho u = \rho (V-p(\rho))$, we deduce that 
  \begin{eqnarray*}
    & & \hspace{-1cm} g = \rho (1- \frac{p(\rho)}{V}), \quad s = \rho \frac{p(\rho)}{V}  . 
  \end{eqnarray*}
  Not surprisingly, the offset velocity scaled by the particle velocity is nothing but the proportion of steady particles 
  and it is completely determined by the total density $\rho$.

  We deduce from (\ref{RARPed_rhou=Vg}) that system (\ref{RARPed_n}), (\ref{RARPed_u}) can be rewritten in the form: 
  \begin{eqnarray}
    & & \hspace{-1cm} \partial_t ( g + s ) + \partial_x (V g) = 0 , \label{RARPed_n2} \\
    & & \hspace{-1cm} \partial_t (V g )  + \partial_x (V^2 g ) = V \pi'(\rho) \partial_x g , \label{RARPed_u2}
  \end{eqnarray}
  Dividing (\ref{RARPed_u2}) by $V$ and subtracting to (\ref{RARPed_n2}), we find: 
  \begin{eqnarray*}
    & & \hspace{-1cm} \partial_t g  + \partial_x (V g) = \pi'(\rho) \partial_x g , \\
    & & \hspace{-1cm} \partial_t s   = - \pi'(\rho) \partial_x g . 
  \end{eqnarray*}
  Thus, the term $\pi'(\rho) \partial_x g$ represents the algebraic transfer rate from immobile to moving particles, while $- \pi'(\rho) \partial_x g$ represents the opposite transfer. Therefore, this model assumes that the pedestrians decide to stop or become mobile again, based not only on local observation of the surrounding density, but on the observation of their gradients. More precisely, 
  keeping in mind that $\pi'(\rho)$ and $V$ have the same sign, the transfer rate from the immobile to moving state is positive if the moving particle density increases in the downstream direction, indicating a lower congestion. Symmetrically, the transfer rate from the moving to immobile state increases if the moving particle density decreases in the downstream direction, indicating an increase of congestion. These evaluations of the variation of the moving particle density derivative are weighted by increasing functions of the density, meaning that the reactions of the pedestrians to their environment are faster if the density is large. 
\end{remark}

We now turn to the introduction of the density constraint in the 1W-AR or 1W-CAR models.

\subsection{Introduction of the maximal density constraint in the 1W-AR model}
\label{subsec_AR_density_constraint}

The maximal density constraint (also referred to below as the congestion constraint) is implemented in the expression of the velocity offset or pressure $p$. Two ways to achieve this goal are proposed. 

In the first one, $p$ is a smooth function of the particle density which blows-up at the approach of the maximal allowed density $\rho^*$. 

In the second one, congestion results in an incompressibility constraint which produces non-local effects with infinite speed of propagation of information. In congested regions, the pressure is no longer a function of the density but becomes implicitly determined by the incompressibility constraint. The transition from uncongested to congested regions is abrupt and appears as a kind of phase transition. This second approach can be realized as an asymptotic limit of the first approach where compression waves (or acoustic waves by analogy with gas dynamics) propagate at larger and larger speeds (so-called low Mach-number limit). 

Below, we successively discuss these two strategies. Then, we specifically consider the introduction of the congestion constraint within the 1W-CAR model.

\subsubsection{Congestion model with smooth transitions between uncongested and congested regions}
\label{subsubsec_AR_density_constraint_smooth}

To implement the congestion constraint, we will highly rely on previous work \cite{BDDR,BDLMRR,Deg_Del}, where this constraint has been implemented in the 1W-AR model. We take a convex function $p(\rho)$ such that $p(0) = 0$, $p'(0) \geq 0$ and $p(\rho) \to \infty$ as $\rho \to \rho^*$. More explicitly, we can choose for instance for the pressure:
\begin{eqnarray}
  & & \hspace{-1cm} p(\rho) = p^\varepsilon (\rho) = P(\rho) + Q^\varepsilon(\rho) , \label{AR_p_blow} \\
  & & \hspace{-1cm} P(\rho) = M \rho^{m}, \quad m > 1,  \label{AR_p_blow2} \\
  & & \hspace{-1cm}  Q^\varepsilon(\rho) = \frac{\varepsilon}{  \left( \frac{1}{\rho} - \frac{1}{\rho^*} \right)^\gamma }  , \quad \gamma > 1.  \label{AR_p_blow3}
\end{eqnarray}
$P(\rho)$ is the background pressure of the pedestrians in the absence of congestion (and is taken in the form of an isentropic gas dynamics equation of state). $Q^\varepsilon$ is a correction which turns on when the density is close to congestion (i.e. $\varepsilon \ll 1$ is a small quantity), and modifies the background pressure to have it match the congestion condition $p(\rho) \to \infty$ as $\rho \to \rho^*$. 

Indeed, as long as $\rho-\rho_*$ is not too small, the denominator in (\ref{AR_p_blow3}) is finite and $Q^\varepsilon(\rho)$ is of order $\varepsilon$. Thus the pressure $p$ is dominated by the $P$ term. However, a crossover occurs when
\begin{displaymath}
  \left( \frac{1}{\rho} - \frac{1}{\rho^*} \right)^\gamma \sim \varepsilon, 
\end{displaymath}
i.e. when 
\begin{equation}
  \rho^* - \rho \sim \rho \rho^* \varepsilon^{1/\gamma}. \label{crossover2}
\end{equation}
Thus in a density range near $\rho^*$ which scales as $\varepsilon^{1/\gamma}$,
the correction $Q^\varepsilon(\rho)$ becomes of order unity. 

\begin{figure}
\begin{center}
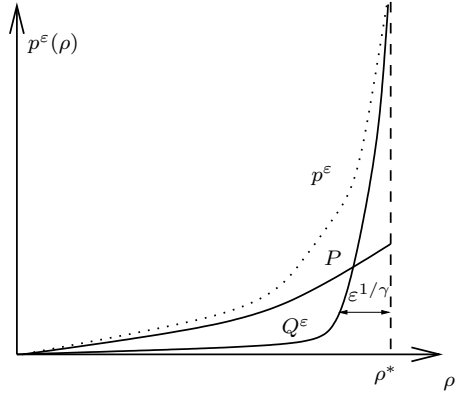
\caption{Schematic representation of
$Q^\varepsilon$, $P$, and $p^\varepsilon= P + Q^\varepsilon$, as a function of $\rho$.}
\label{fig_p}
\end{center}
\end{figure}

This is represented schematically on Figure \ref{fig_p}.
Note that the precise shape of the term 
$\left( \frac{1}{\rho} - \frac{1}{\rho^*} \right)^\gamma$
is not important, as it does not contribute to the pressure law, except in a narrow region close to congestion.
The chosen expression ensures that $Q^\varepsilon(\rho=0)=0$,
and that it becomes significant
in the vicinity of $\rho^*$ only. Note also that 
$Q^\varepsilon$ is an increasing function of $\rho$, in order
to keep the problem hyperbolic.

The pressure singularity at $\rho = \rho^*$ ensures that the congestion density $\rho^*$ cannot be exceeded. Indeed, let us consider a closed system (e.g. the system is posed on an interval $[a,b]$ with periodic boundary conditions) for simplicity.  Let $u_0$ and $w_0$ be the initial conditions and suppose that they satisfy
\begin{eqnarray*}
  & & \hspace{-1cm} 0 \leq u_{\mbox{\scriptsize m}} \leq u_0 \leq u_{\mbox{\scriptsize M}}, \quad  0 \leq w_{\mbox{\scriptsize m}} \leq w_0 \leq w_{\mbox{\scriptsize M}}, 
\end{eqnarray*}
for some constants $u_{\mbox{\scriptsize m}}$, $u_{\mbox{\scriptsize M}}$,
$w_{\mbox{\scriptsize m}}$, $w_{\mbox{\scriptsize M}}$. Then, \cite{AR} notices that, at
any time, $u$ and $w$ satisfy the same estimates:
\begin{eqnarray}
  & & \hspace{-1.1cm} 0 \leq u_{\mbox{\scriptsize m}} \leq u(x, t) \leq u_{\mbox{\scriptsize M}}, \quad  0 \leq w_{\mbox{\scriptsize m}} \leq w(x,t) \leq w_{\mbox{\scriptsize M}}, \quad \forall (x,t) \in  [a,b] \times {\mathbb R}_+.  \label{estim}
\end{eqnarray}
In other words, this estimate defines an invariant region of the system. It follows from the fact that, $u$ and $w$ being the two Riemann invariants, they are transported by the characteristic fields (see eqs.  (\ref{AR_u_lag}), (\ref{AR_w_lag})) and therefore, satisfy the maximum principle. From (\ref{estim}), we deduce that $w - u = p(\rho) \leq w_{\mbox{\scriptsize M}} - u_{\mbox{\scriptsize m}}$, and we also have that $p(\rho) \geq 0$ at all times.  Let $p^{-1}$ be the inverse function of $p$. Since $p$ maps $[0,\rho^*)$ increasingly to ${\mathbb R}_+$, then $p^{-1}$ maps increasingly ${\mathbb R}_+$ onto $[0,\rho^*)$, from which the estimate $\rho \leq p^{-1} (w_{\mbox{\scriptsize M}} - u_{\mbox{\scriptsize m}}) < \rho^*$ follows. This indeed shows that the constraint $\rho < \rho^*$ is satisfied at all times. From the estimate (\ref{estim}), we also see that $u$ cannot become negative, so that the estimate $w \geq p(\rho)$ is also satisfied at all times.

With this $\varepsilon$-dependent pressure, the 1W-AR model becomes a perturbation problem, written as follows: 
\begin{eqnarray}
  & & \hspace{-1cm} \partial_t \rho^\varepsilon  + \partial_x (\rho^\varepsilon u^\varepsilon) = 0 , \label{EAR_n} \\
  & & \hspace{-1cm} \partial_t (\rho^\varepsilon w^\varepsilon)  + \partial_x (\rho^\varepsilon w^\varepsilon u^\varepsilon) = 0 , \label{EAR_u} \\
  & & \hspace{-1cm} w^\varepsilon= u^\varepsilon + p^\varepsilon(\rho^\varepsilon). \label{EAR_w} 
\end{eqnarray}
The next section investigates the formal $\varepsilon \to 0$ limit.

\subsubsection{Congestion model with abrupt transitions between uncongested and congested regions}
\label{subsubsec_AR_density_constraint_incompressibility}

In the limit $\varepsilon \to 0$, the uncongested motion remains unperturbed until the density hits the exact value $\rho^*$. Once this happens, congestion suddenly turns on and modifies the dynamics abruptly. In the uncongested regions, the flow is compressible ; it becomes incompressible at the congestion density $\rho^*$. Therefore, in the limit $\varepsilon \to 0$, the abrupt transition from uncongested motion (when $\rho < \rho^*$) to congested motion (when $\rho = \rho^*$) corresponds to the crossing of a phase transition between a compressible to an incompressible flow regime. 

In the limit $\varepsilon \to 0$, the arguments of \cite{BDDR,BDLMRR,Deg_Del} can be easily adapted. Suppose that $\rho^\varepsilon \to \rho < \rho^*$. In this case, $Q^\varepsilon (\rho^\varepsilon) \to 0$ and we recover an 1W-AR model associated to the pressure $P(\rho)$:   
\begin{eqnarray}
  & & \hspace{-1cm} \partial_t \rho^0  + \partial_x (\rho^0 u^0) = 0 , \label{0AR_n_NC} \\
  & & \hspace{-1cm} \partial_t (\rho^0 w^0)  + \partial_x (\rho^0 w^0 u^0) = 0 , \label{0AR_u_NC} \\
  & & \hspace{-1cm} w^0= u^0 + P(\rho^0). \label{0AR_w_NC} 
\end{eqnarray}
If on the other hand, $\rho^\varepsilon \to \rho^*$, then $Q^\varepsilon (\rho^\varepsilon) \to \bar Q$ with $0 \leq \bar Q \leq w_{\mbox{\scriptsize M}}$. Therefore, the total pressure is such that $p^\varepsilon (\rho^\varepsilon)  \to \bar p$ with $P(\rho^*) \leq \bar p $. In this case, the model becomes incompressible: 
\begin{eqnarray}
  & & \hspace{-1cm} \partial_x  u^0 = 0 , \label{0AR_n_C} \\
  & & \hspace{-1cm} \partial_t w^0  + u^0 \partial_x  w^0  = 0 , \label{0AR_u_C} \\
  & & \hspace{-1cm} w^0= u^0 + \bar p, \quad \mbox{ with } \quad P(\rho^*) \leq \bar p .  \label{0AR_w_C} 
\end{eqnarray}
Note that in this congested region, the density does not vary (it is equal to $\rho^*$)
and cannot determine the pressure anymore. Indeed, the functional relation between the density and the pressure is broken and $\bar p$ may be varying with $x$ even though $\rho$ does not. The spatial variations of $\bar p$ compensate
exactly (through (\ref{0AR_w_C})) the variations of $w^0$, in such a way that all the
pedestrians, whatever their desired velocity is, move at the same speed in the congestion
region.

This can also be seen when taking the limit $\varepsilon \to 0$ in (\ref{AR_u_lag}). Indeed, if $\rho^\varepsilon \to \rho^*$ with $p^\varepsilon (\rho^\varepsilon) ( = w^\varepsilon - u^\varepsilon )$ staying finite, then $\rho^\varepsilon - \rho^* = O(\varepsilon^{1/\gamma})$ (see (\ref{crossover2})) and $d p^\varepsilon / d \rho \sim \varepsilon^{-1/\gamma} \to \infty$. Therefore, in the congested regime, the derivative of the pressure with respect to the density becomes infinite. Inserting this in (\ref{AR_u_lag}) shows that $\partial_x u^\varepsilon \to 0$. This ensures that all
the pedestrians move at the same speed. Simultaneously, this blocks any further increase of the density, which cannot become larger than $\rho^*$. Indeed, the mass conservation equation (\ref{AR_n}) tells us that
\begin{eqnarray*}
  & & \hspace{-1cm} \frac{d}{dt} \rho^\varepsilon  = \partial_t \rho^\varepsilon  + u^\varepsilon \partial_x  \rho^\varepsilon  = - \rho^\varepsilon \partial_x u^\varepsilon ,  
\end{eqnarray*}
and consequently, if $\partial_x u^\varepsilon \to 0$, any further increase of the density is impeded.

In the general case, we expect that the two limit regimes coexist.  The congested region
may appear anywhere in the flow, depending on the initial conditions. Congestion regions
must be connected to uncongested regions by interface conditions. Across these interfaces,
$\rho$ and $\rho w$, which are conserved quantities obey the Rankine-Hugoniot relations.
The quantity $w$, which is thought of as the (locally averaged) pedestrians' desired velocity is modified across the interfaces through these relations. However, the bounds (\ref{estim}) are preserved (see \cite{AR}).

Connecting congested and uncongested regions is a delicate problem which has been investigated in \cite{BDDR} by a careful inspection of Riemann problem solutions. Specifically, \cite{BDDR} treats the special case $M=0$ in (\ref{AR_p_blow})-(\ref{AR_p_blow3}). The present choice of the pressure (\ref{AR_p_blow})-(\ref{AR_p_blow3}) is slightly different: in the limit $\varepsilon \to 0$, it produces a non-zero pressure in the uncongested region, while \cite{BDDR} considers that uncongested regions are pressureless in this limit. Pressureless gas dynamics develops some unpleasant features (such as the occurrence of vacuum, weak instabilities, and so on). Keeping a non-zero pressure in the uncongested region in the limit $\varepsilon \to 0$ allows to bypass some of these problems and represents an improvement over \cite{BDDR}. Of course, the precise choice of $m$ and $M$ must be fitted against experimental data. 

We do not attempt to derive interface conditions between uncongested and congested regions for the present choice of the pressure. Indeed, the perturbation problem (\ref{EAR_n})-(\ref{EAR_w}), even with a small value of $\varepsilon$ is easier to treat numerically than the connection problem between the two models (\ref{0AR_n_NC})-(\ref{0AR_w_NC}) and (\ref{0AR_n_C})-(\ref{0AR_w_C}). Therefore, we will not regard the limit model as a numerically effective one, but rather, as a theoretical limit which provides some useful insight. Still, the numerical treatment of the perturbation problem requires some care. Of particular importance is the development of Asymptotic-Preserving schemes, i.e. of schemes that are able to capture the correct asymptotic limit when $\varepsilon \to 0$. This is not an easy problem because of the blow up of the pressure near $\rho^*$. Indeed, due to the blow up of the characteristic speed in (\ref{AR_u_lag}), the CFL stability condition of a classical explicit shock-capturing method leads to a time-step constraint of the type $\Delta t =0( \varepsilon^{1/\gamma} ) \to 0$ as $\varepsilon \to 0$. For this reason, classical explicit shock-capturing methods cannot be used to explore the congestion constraint when $\varepsilon \to 0$ and Asymptotic-Preserving schemes are needed. 

Another reason for considering the perturbation problem (\ref{EAR_n})-(\ref{EAR_w}) instead of the limit model 
is that the congestion may appear gradually rather than like an abrupt phase transition from compressible to incompressible motion. In particular, for large pedestrian concentrations, some erratic motions occur (this is referred to as crowd turbulence) and might be modeled by a suitable (may be different) choice of the perturbation pressure $Q^\varepsilon$.

\subsubsection{Introduction of the congestion constraint in the constant desired velocity 1W-CAR model}
\label{subsubsec_TAR_density_constraint}

\mbox{}

\noindent {\it (i) Congestion model with smooth transitions.} The smooth pressure relations (\ref{AR_p_blow})-(\ref{AR_p_blow3}) can be used for the 1W-CAR model. Because $\rho$ now satisfies a convection equation: 
\begin{eqnarray*}
  & & \hspace{-1cm} \partial_t \rho + (V - (\rho p )'(\rho)) \partial_x \rho =0,  
\end{eqnarray*}
the initial bounds are preserved. Indeed, suppose that 
\begin{eqnarray*}
  & & \hspace{-1cm} 0 \leq \rho_{\mbox{\scriptsize m}} \leq \rho_0 \leq \rho_{\mbox{\scriptsize M}} < \rho^*, 
\end{eqnarray*}
for some constants $\rho_{\mbox{\scriptsize m}}$, $\rho_{\mbox{\scriptsize M}}$, then, at any time, $\rho$ satisfy the same estimates. 
\begin{eqnarray*}
  & & \hspace{-1cm} 0 \leq \rho_{\mbox{\scriptsize m}} \leq \rho(x, t) \leq \rho_{\mbox{\scriptsize M}} < \rho^*, \quad \forall (x,t) \in  [a,b] \times {\mathbb R}_+.  
\end{eqnarray*}
In this way, the constraint $0 \leq \rho \leq \rho^*$ is always satisfied. However, the fact that the bounds on the density are preserved by the dynamics can be viewed as unrealistic. In real pedestrian traffic, strips of congested and uncongested traffic spontaneously emerge from rather space homogeneous initial conditions. The generation of new maximal and minimal bounds is an important feature of real traffic systems which is not well taken into account in the 1W-CAR model and more generally, in LWR models.

\medskip

\noindent {\it (ii) Congestion model with abrupt transitions.} If the limit $\varepsilon \to 0$ is considered, and if the upper bound $\rho_{\mbox{\scriptsize M}} = \rho_{\mbox{\scriptsize M}}^\varepsilon$ depends on $\varepsilon$ and is such that $\rho_{\mbox{\scriptsize M}}^\varepsilon \to \rho^*$, then, some congestion regions can occur. The limit model in the uncongested region does not change, and is given by the single conservation relation (\ref{ARPed_n}) with the pressure $p(\rho^0)= P(\rho^0)$. In the congested region, we have $\rho^0 = \rho^*$, which implies $\partial_x u^0 = 0$. In terms of the moving and steady pedestrian densities, the congested regime means that 
\begin{eqnarray*}
  & & \hspace{-1cm} \partial_x g^0 = 0, \qquad s^0 = \rho^* - g^0  , 
\end{eqnarray*}
i.e. both the steady and moving pedestrian densities are uniform in the congested region.

\section{Two-way one-lane traffic model}
\label{1lane}

\subsection{An Aw-Rascle model for two-way one-lane pedestrian traffic}
\label{subsec_2AR_model}

The extension of the 1W-AR model to 2-way traffic, denoted by 2W-AR model, may seem rather
easy, the 2-way traffic is written as a system of two 1-way models. However, we will see
that the mathematical properties of the 2-way models are rather different from their
one-way counterpart.

\begin{definition} {\bf (2W-AR model) }
  Let $\rho_\pm$ the density of pedestrians, $u_\pm$ their velocity, $w_\pm$ their desired
  velocity and $p$ the pressure, with an index $+$ for the right-going pedestrians and $-$
  for the left-going ones. The 2W-AR model for 2-way traffic is written:
  \begin{eqnarray*}
    & & \hspace{-1cm} \partial_t \rho_+ + \partial_x (\rho_+ u_+) = 0 , \label{2AR_n+} \\
    & & \hspace{-1cm} \partial_t \rho_- + \partial_x (\rho_- u_-) = 0 , \label{2AR_n-} \\
    & & \hspace{-1cm} \partial_t (\rho_+ w_+)  + \partial_x (\rho_+ w_+ u_+) = 0 , \label{2AR_u+} \\
    & & \hspace{-1cm} \partial_t (\rho_- w_-)  + \partial_x (\rho_- w_- u_-) = 0 , \label{2AR_u-} \\
    & & \hspace{-1cm} w_+= u_+ + p(\rho_+, \rho_-),  \label{2AR_w+} \\
    & & \hspace{-1cm} w_-= -u_- + p(\rho_-, \rho_+) . \label{2AR_w-} 
  \end{eqnarray*}
\end{definition}

The coupling of the two flows of pedestrians in the 2W-AR model is through the
prescription of the pressures which are functions of the densities of the two
species $\rho_+$ and $\rho_-$. Our conventions are that the desired velocities $w_\pm$ and the
velocity offsets $p(\rho_\pm, \rho_\mp)$ are magnitudes, and as such, are positive
quantities. The actual velocities $u_\pm$ are signed quantities: $u_+ >0$ for right-going
pedestrians and $u_-<0$ for left-going pedestrians. These conventions explain the
different signs in factor of the velocities for (\ref{2AR_w+}) and
(\ref{2AR_w-}). However, we do not exclude that, in particularly congested conditions, the
right-going pedestrians may have to go backwards (i.e. to the left) or vice-versa, the
left-going pedestrians have to go to the right. Therefore, we do not make any a priori
assumption on the sign of $u_\pm$. For obvious symmetry reasons, the same pressure
function is used for the two particles, with reversed arguments. The function $p$ is
increasing with respect to both arguments since the velocity offset of one of the species
increases when the density of either species increases.

Some of the properties of the 1W-AR system extend to the 2W-AR one. For instance, the desired velocities are Lagrangian variables, as they satisfy: 
\begin{eqnarray*}
  & & \hspace{-1cm} \partial_t w_+  + u_+ \partial_x w_+ = 0  , \label{2AR_w+_lag} \\
  & & \hspace{-1cm} \partial_t w_-  + u_- \partial_x w_- = 0  . \label{2AR_w-_lag} 
\end{eqnarray*}
Unfortunately, the velocities $u_+$ and $u_-$ do not constitute Riemann invariants any longer because of the coupling induced by the dependence of $p$ upon $\rho_+$ and $\rho_-$.
For this reason initial bounds on $u_+$ and $u_-$ are not preserved by the flow, as they were in the case of the 1W-AR model. Since the velocity offsets $p(\rho_+, \rho_-)$ and $p(\rho_-, \rho_+)$ are not bounded a priori, the velocities $u_+$ and $u_-$ can reverse sign when the velocity offsets are large. This is expected to reflect the fact that a dense crowd moving in one direction may force isolated pedestrians going the other way to move backwards. Of course, such a situation is only expected in close to congestion regimes. 

Nonetheless, the evolution of the pedestrian fluxes reflects the same phenomenology as in the one-way case, namely that pedestrians react to the Lagrangian derivative of the pressure, as shown by the following eqs. (which are the 2-way equivalents of eq. (\ref{AR_rho_u})):
\begin{eqnarray*}
  & & \hspace{-1cm} \partial_t (\rho_+ u_+)  + \partial_x (\rho_+ u_+ u_+ ) =  - \rho_+ \, 
  \left( \frac{d}{dt} \right)_+ [ p(\rho_+,\rho_-) ]  , \label{2AR_rho_u+} \\
  & & \hspace{-1cm} \partial_t (\rho_- u_-)  + \partial_x (\rho_- u_- u_- ) =   \rho_- \, 
  \left( \frac{d}{dt} \right)_- [ p(\rho_-,\rho_+) ]  , \label{2AR_rho_u-} 
\end{eqnarray*}
where the material derivatives $(d/dt)_{\pm} = \partial_t + u_{\pm} \partial_x$ depend on what type of  particles is concerned.  These equations can also be put in the form
(equivalent to (\ref{AR_rho_u_2}) for the 1W-AR model): 
\begin{eqnarray}
  & & \hspace{-1cm}\partial_t (\rho_+ u_+)  + \partial_x (\rho_+ u_+ w_+ ) =  \left[ p(\rho_+,\rho_-) + \rho_+ \left.\partial_1 p\right|_{(\rho_+,\rho_-)} \right]  \partial_x (\rho_+ u_+) + \nonumber \\
  & & \hspace{6cm} + \rho_+ \left.\partial_2 p\right|_{(\rho_+,\rho_-)}   \partial_x (\rho_- u_-) , \label{2AR_rho_u+_2} \\
  & & \hspace{-1cm} \partial_t (\rho_- u_-)  - \partial_x (\rho_- u_- w_- ) =  - \left[p(\rho_-,\rho_+) + \rho_- \left.\partial_1 p\right|_{(\rho_-,\rho_+)} \right]  \partial_x (\rho_- u_-)  \nonumber \\
  & & \hspace{6cm} - \rho_- \left.\partial_2 p\right|_{(\rho_-,\rho_+)}   \partial_x (\rho_+ u_+) , \label{2AR_rho_u-_2} 
\end{eqnarray}
where we denote by $\partial_1 p$ and $\partial_2 p$ the derivatives of the function $p$ with respect to its first and second arguments respectively.
This form of the equations will be used below for the derivation of the Constant Desired Velocity model.

\medskip
The 2W-AR model is not always hyperbolic. Before stating the result, we introduce some notations. 
We define:
\begin{eqnarray}
  & & \hspace{-1cm} 
  c_{++} = \partial_1 p(\rho_+,\rho_-), \quad \quad
  c_{+-} = \partial_2 p(\rho_+,\rho_-),  \label{eq:speed_def1} \\
  & & \hspace{-1cm} 
  c_{-+} = \partial_2 p(\rho_-,\rho_+), \quad \quad
  c_{--} = \partial_1 p(\rho_-,\rho_+). \label{eq:speed_def2}
\end{eqnarray}
We assume that $p$ is increasing with respect to both arguments, which implies that all quantities defined by (\ref{eq:speed_def1}), (\ref{eq:speed_def2}) are non-negative. This assumption simply means that the pedestrian speed is reduced if the densities of either categories of pedestrians increase.  
For a given state $(\rho_+,w_+,\rho_-,w_-)$, the fluid velocities are given by: 
\begin{displaymath}
  u_+ = w_+ - p(\rho_+,\rho_-), \quad u_- = - w_- + p(\rho_-, \rho_+). 
\end{displaymath}
We also define the following velocities 
\begin{displaymath}
  c_{u_+} = u_+ - \rho_+ c_{++}, \quad c_{u_-} = u_- + \rho_- c_{--} . 
  \label{eq:vel_waves}
\end{displaymath}
These are the characteristic speeds (\ref{AR_u_vel}) of the 1W-AR system. Specifically, $c_{u_+}$ is the wave at which information about velocity would propagate in a system of right-going pedestrians without coupling with the left-going ones. A similar explanation holds symmetrically for $c_{u_-}$. 

We now have the following theorem, the proof of which is elementary and left to the reader. 

\begin{theorem}
  The 2W-AR system is hyperbolic about the state $(\rho_+,w_+,\rho_-,w_-)$ if and only if the following condition holds true: 
  \begin{equation}
    \Delta := (c_{u_+} - c_{u_-})^2 - 4 \rho_+ \rho_- c_{+-} c_{-+} \geq 0.
    \label{eq:hyp_cond}
  \end{equation}
  The quantities $u_\pm$ are two characteristics velocities of the system. If condition (\ref{eq:hyp_cond}) is satisfied, the two other characteristic velocities are 
  \begin{equation}
    \lambda_{\pm} = \frac{1}{2} \left[  c_{u_+} + c_{u_-} \pm \sqrt \Delta \right]. 
    \label{eq:char_vel_2}
  \end{equation}
  \label{thm_hyp_2WAR}
\end{theorem}

\noindent
Non-hyperbolicity occurs when the two characteristic velocities $c_{u_+}$ and $c_{u_-}$ of
the uncoupled systems are close to each other. In this case, the first term of
$\Delta$ is close to zero and does not compensate for the second term, which is
negative. These conditions happen in particular when  both velocities $c_{u_+}$ and
$c_{u_-}$ are close to zero, which corresponds to the densities where the fluxes $\rho_+
u_+,$ $\rho_- u_-$ are maximal as functions of the densities $\rho_+$, $\rho_-$
respectively. In particular, in the one-way case with constant speed of figure \ref{Fig_LWR}, this would
correspond to the point $\rho_{\text{max}}$. These conditions correspond to the onset of
congestion. Therefore, instabilities linked to the non-hyperbolic character of the model
will develop in conditions close to congestion.

The occurrence of regions of non-hyperbolicity is not entirely surprising. The instability of two counter-propagating flows is a common phenomenon in fluid mechanics. In plasma physics, the instability of two counter-propagating streams of charged particles is well known under the two-stream instability. The situation here is extremely similar, in spite of the different nature of the interactions (which are mediated by the long-range Coulomb force in the plasma case). 

The occurrence of a non-hyperbolic region is often viewed as detrimental, because in this region, the model is unstable. On the other hand, self-organization phenomena like lane formation or the onset of crowd turbulence cannot be described by an everywhere stable model. For instance, morphogenesis is explained by the occurrence of the Turing instability in systems of diffusion equations. Here, diffusion is not taken into account and the instability originates from a different phenomenon. However, in practice, some small but non-zero diffusion always exists. This diffusion damps the small scale structures but keeps the large scale structures growing. The typical size of the observed structures can be linked to the threshold wave-number below which instability occurs. 

Numerical simulations to be presented in a forthcoming work will allow us to determine whether the phenomena which are observed in dense crowds may be explained by this type of instability. In section \ref{sec_math}, a stability analysis of a diffusive two-way LWR model will provide more quantitative support to these concepts.

\subsection{The constant desired velocity Aw-Rascle model for two-way one-lane pedestrian traffic}
\label{subsec_2TAR_model}

To construct the two-way constant desired velocity Aw-Rascle model (2W-CAR) for two-way one-lane pedestrian traffic, we must set 
\begin{eqnarray}
  & & \hspace{-1cm} w_+ = w_-= V  , \label{2ARPed_w=v}
\end{eqnarray}
and 
\begin{eqnarray}
  & & \hspace{-1cm} u_+ = V  - p(\rho_+,\rho_-)  , \quad u_- = -V  + p(\rho_-,\rho_+)  . \label{2ARPed_u}
\end{eqnarray}
This leads to the following model:

\begin{definition} {\bf (2W-CAR model)} Let $\rho_+$ and $\rho_-$ the densities of
  pedestrians moving to the right and to the left respectively, $V$ the (constant) desired velocity of
  pedestrians and $p$ the pressure term. The 2W-CAR model is written:
  \begin{eqnarray}
    & & \hspace{-1cm} \partial_t \rho_+ + \partial_x (\rho_+ (V - p(\rho_+,\rho_-)) ) = 0 , \label{2ARPed_n+} \\
    & & \hspace{-1cm} \partial_t \rho_- - \partial_x (\rho_- (V - p(\rho_-,\rho_+)) ) = 0 . \label{2ARPed_n-} 
  \end{eqnarray}
\end{definition}

These are two first-order models coupled by a velocity offset which depends on the two densities.

We can find the same interpretation of this model in terms of moving and steady particles as in the one-way model case. Using (\ref{2AR_rho_u+_2}), (\ref{2AR_rho_u-_2}) and (\ref{2ARPed_w=v}), we can write: 
\begin{eqnarray}
  & & \hspace{-1cm} \partial_t \rho_+ + \partial_x (\rho_+ u_+) = 0 , \label{R2TAR_n+} \\
  & & \hspace{-1cm} \partial_t \rho_- + \partial_x (\rho_- u_-) = 0 , \label{R2TAR_n-} \\
  & & \hspace{-1cm}\partial_t (\rho_+ u_+)  + \partial_x (\rho_+ u_+ V ) = 
  \left[ p(\rho_+,\rho_-) + \rho_+ \left.\partial_1 p\right|_{(\rho_+,\rho_-)} \right]  \partial_x (\rho_+ u_+) + \nonumber \\
  & & \hspace{6cm} + \rho_+ \left.\partial_2 p\right|_{(\rho_+,\rho_-)}   \partial_x (\rho_- u_-) , \label{R2TAR_rho_u+} \\
  & & \hspace{-1cm} \partial_t (\rho_- u_-)  - \partial_x (\rho_- u_- V ) =  - \left[p(\rho_-,\rho_+) + \rho_- \left.\partial_1 p\right|_{(\rho_-,\rho_+)} \right]  \partial_x (\rho_- u_-)  \nonumber \\
  & & \hspace{6cm} - \rho_- \left.\partial_2 p\right|_{(\rho_-,\rho_+)}   \partial_x (\rho_+ u_+) . \label{R2TAR_rho_u-} 
\end{eqnarray}
Conversely, if $\rho_+$, $u_+$, $\rho_-$, $u_-$ are solutions of this model, using the same method as in the one-way case, we easily find that: 
\begin{eqnarray*}
  & & \hspace{-1cm} \partial_t (\rho_\pm (p \pm u_\pm - V) ) = 0  . \label{R2ARPed_const}
\end{eqnarray*}
Therefore, if (\ref{2ARPed_u}) is satisfied initially, it is satisfied at all times and we recover (\ref{2ARPed_n+}), (\ref{2ARPed_n-}). 

Now, we denote by $g_\pm(x,t)$ the density of the moving particles and by $s_\pm(x,t)$ that of the steady particles with a $+$ (respectively a $-$) indicating the right-going (respectively left-going) pedestrians. Although steady, the pedestrians have a desired motion either to the right or to the left, and we need to keep track of these intended directions of motions. We have
\begin{eqnarray*}
  & & \hspace{-1cm} \rho_\pm = g_\pm + s_\pm  \quad \mbox{ and } \quad \rho_\pm u_\pm = \pm V g_\pm  . \label{R2ARPed_rhou=Vg}
\end{eqnarray*}
We deduce that 
\begin{eqnarray*}
  & & \hspace{-1cm} \frac{s_+}{\rho_+}  = \frac{p(\rho_+,\rho_-)}{V}, \quad \quad  \frac{s_-}{\rho_-}  = \frac{p(\rho_-,\rho_+)}{V}. \label{R2ARPed_s_pm}
\end{eqnarray*} 
Therefore, the offset velocities $p(\rho_+,\rho_-)$ and $p(\rho_-,\rho_+)$ scaled by the particle velocity $V$ represent the proportions of the steady particles $s^+/\rho_+$  and $s^-/\rho_-$ respectively. Now, we can rewrite  (\ref{R2TAR_n+})-(\ref{R2TAR_rho_u-}) as follows:
\begin{eqnarray*}
  & & \hspace{-1cm} \partial_t (g_+ + s_+) + \partial_x (V g_+) = 0 , \label{R2TAR_n+2} \\
  & & \hspace{-1cm} \partial_t (g_- + s_-) - \partial_x (V g_-) = 0 , \label{R2TAR_n-2} \\
  & & \hspace{-1cm}\partial_t (V g_+)  + \partial_x (V^2 g_+ ) =  \left[ p(\rho_+,\rho_-) + \rho_+ \left.\partial_1 p\right|_{(\rho_+,\rho_-)} \right]  \partial_x (V g_+)  \nonumber \\
  & & \hspace{6cm} - \rho_+ \left.\partial_2 p\right|_{(\rho_+,\rho_-)}   \partial_x (V g_-) , \label{R2TAR_rho_u+2} \\
  & & \hspace{-1cm}  \partial_t (V g_-)  - \partial_x (V^2 g_- ) =  - \left[p(\rho_-,\rho_+) + \rho_- \left.\partial_1 p\right|_{(\rho_-,\rho_+)} \right]  \partial_x (V g_-)  + \nonumber \\
  & & \hspace{6cm} + \rho_- \left.\partial_2 p\right|_{(\rho_-,\rho_+)}   \partial_x (V g_+) . \label{R2TAR_rho_u-2} 
\end{eqnarray*}
By simple linear combinations, this system is equivalent to
\begin{eqnarray*}
  & & \hspace{-1cm}\partial_t g_+  + \partial_x (V g_+ ) =  \left[ p(\rho_+,\rho_-) + \rho_+ \left.\partial_1 p\right|_{(\rho_+,\rho_-)} \right]  \partial_x g_+  \nonumber \\
  & & \hspace{6cm} - \rho_+ \left.\partial_2 p\right|_{(\rho_+,\rho_-)}   \partial_x g_- , \label{R2TAR_rho_g+} \\
  & & \hspace{-1cm}  \partial_t g_-  - \partial_x (V g_- ) =  - \left[p(\rho_-,\rho_+) + \rho_- \left.\partial_1 p\right|_{(\rho_-,\rho_+)} \right]  \partial_x g_-  + \nonumber \\
  & & \hspace{6cm} + \rho_- \left.\partial_2 p\right|_{(\rho_-,\rho_+)}   \partial_x g_+ , \label{R2TAR_rho_g-} \\
  & & \hspace{-1cm}\partial_t s_+  =  - \left[ p(\rho_+,\rho_-) + \rho_+ \left.\partial_1 p\right|_{(\rho_+,\rho_-)} \right]  \partial_x g_+ + \rho_+ \left.\partial_2 p\right|_{(\rho_+,\rho_-)}   \partial_x g_- , \label{R2TAR_rho_s+} \\
  & & \hspace{-1cm}  \partial_t s_-   =   \left[p|_{(\rho_-,\rho_+)} + \rho_- \partial_1 p|_{(\rho_-,\rho_+)} \right]  \partial_x g_- - \rho_- \partial_2 p|_{(\rho_-,\rho_+)}   \partial_x g_+ . \label{R2TAR_rho_s-} 
\end{eqnarray*}

Like in the one-way model, we find that the transition rates from the steady to moving states or vice-versa depend on the derivatives of the concentrations of moving pedestrians. Now, both the left and right going pedestrian total densities appear in the expressions of the transitions rates for either species. This is due to the coupling through the pressure term, which depends on both densities.

\medskip
Like the 2W-AR model, the 2W-CAR model is not always hyperbolic. Using the same notations as in the previous section, we have the:

\begin{theorem}
  The 2W-CAR system is hyperbolic about the state $(\rho_+,\rho_-)$ if and only if condition (\ref{eq:hyp_cond}) is satisfied. In this case, the two characteristic velocities are given by (\ref{eq:char_vel_2}). 
  \label{thm_hyp_2WCAR}
\end{theorem}

\noindent
This can be seen directly from equations
(\ref{2ARPed_n+}) and (\ref{2ARPed_n-}), once they are put under the form
\begin{displaymath}
  \partial_t
  \left(\begin{array}{c} \rho_+\\\rho_-\end{array}\right)
  + \left(\begin{array}{cc} c_{u_+} & -\rho_+ c_{+-}\\
      \rho_- c_{-+} & c_{u_-} \end{array}\right)
  \partial_x \left(\begin{array}{c} \rho_+\\\rho_-\end{array}\right)
  =0 .
\end{displaymath}

We refer to the end of section \ref{subsec_2AR_model} for more comments about this property.

\subsection{Introduction of the congestion constraint in the 2W-AR model}
\label{subsec_2WAR_density_constraint}

\subsubsection{Congestion model with smooth transitions}
\label{subsubsec_2WAR_density_constraint_smooth}

It is difficult to make a prescription for the function $p$. Its expression should be fitted to experimental data. Here we propose a form which allows us to investigate the effects of congestion. We propose: 
\begin{eqnarray}
  & & \hspace{-1cm} p(\rho_+, \rho_-) = p^\varepsilon (\rho_+,\rho_-) = P(\rho) + Q^\varepsilon(\rho_+,\rho_-) , \quad \mbox{ with } \quad \rho = \rho_+ + \rho_- \label{2AR_p_blow} \\
  & & \hspace{-1cm} P(\rho) = M \rho^{m}, \quad m \geq 1,  \label{2AR_p_blow2} \\
  & & \hspace{-1cm}  Q^\varepsilon(\rho_+,\rho_-) = \frac{\varepsilon}{q(\rho_+) \left( \frac{1}{\rho} - \frac{1}{\rho^*} \right)^\gamma}  , \quad \gamma > 1.  \label{2AR_p_blow3}
\end{eqnarray}

The rationale for this formula is as follows. First, in uncongested regime, we expect that the velocity offsets of the right and left going pedestrians are the same, this common offset being a function of the total particle density. Thus, the uncongested flow pressure $P$ given by (\ref{2AR_p_blow2}) is a function of $\rho$ only, and has the same shape as in the one-way case. Congestion occurs when the total density $\rho$ becomes close to $\rho^*$. Therefore, formula (\ref{2AR_p_blow3}) resembles (\ref{AR_p_blow3}), except for the prefactor $q(\rho_+)$. With this choice of the pressure, 
we anticipate that the constraint 
\begin{eqnarray*}
  & & \rho = \rho_+ + \rho_- \leq \rho^* \label{2AR_constraint}
\end{eqnarray*}
will be satisfied everywhere in space and time, like in the one-way case. 

The prefactor $q(\rho_+)$ takes into account the fact that the velocity offset for the majority particle is smaller than that of the minority particle. Therefore, we prescribe $q$ to be an increasing function of $\rho_+$. For further usage, we note the following formula, which follows from eliminating $( ({1}/{\rho}) - ({1}/{\rho^*}) )^\gamma$ between $Q^\varepsilon(\rho_+,\rho_-)$ and $Q^\varepsilon(\rho_-,\rho_+)$: 
\begin{eqnarray}
  & & \hspace{-1cm}  q(\rho_+) \, Q^\varepsilon(\rho_+,\rho_-) = q(\rho_-) \, Q^\varepsilon(\rho_-,\rho_+) .   \label{2AR_p_blow4}
\end{eqnarray}
It is more convenient to express this formula as 
$$ \frac{Q^\varepsilon(\rho_+,\rho_-)}{Q^\varepsilon(\rho_-,\rho_+)} = \frac{q(\rho_-)}{q(\rho_+)}, $$
remembering that $q$ is an increasing function. This formula states that the velocity
offset for the right and left-going particles are inversely proportional to the ratios of
a (function of) the densities. Since $q$ is increasing and taking $\rho_- <
\rho_+$ as an example, we deduce that the velocity offset of the right-going particles
will be less than that of the left-going particles. In other words, the flow of the
majority category of pedestrians is less impeded than that of the minority one. In order
to keep $Q^\varepsilon(\rho_\pm,\rho_\mp)$ small whenever $\rho<\rho^*$, we require that
$q(\rho_\pm) = O(1)$ when $\rho_\pm < \rho^*$ . Physically relevant expressions of 
$q(\rho_\pm)$ can be obtained from real experiments. A possible extension, that we will
not consider here, would be to have different functions $q_+(\rho_+)$ and $q_-(\rho_-)$.
This could model the fact that for example, a crowd heading towards a train platform could
be more pushy than the one going in the opposite direction.

The 2W-AR model with $\varepsilon$-dependent pressure becomes a perturbation problem: 
\begin{eqnarray*}
  & & \hspace{-1cm} \partial_t \rho^\varepsilon_+ + \partial_x (\rho^\varepsilon_+ u^\varepsilon_+) = 0 , \label{E2AR_n+} \\
  & & \hspace{-1cm} \partial_t \rho^\varepsilon_- + \partial_x (\rho^\varepsilon_- u^\varepsilon_-) = 0 , \label{E2AR_n-} \\
  & & \hspace{-1cm} \partial_t (\rho^\varepsilon_+ w^\varepsilon_+)  + \partial_x (\rho^\varepsilon_+ w^\varepsilon_+ u^\varepsilon_+) = 0 , \label{E2AR_u+} \\
  & & \hspace{-1cm} \partial_t (\rho^\varepsilon_- w^\varepsilon_-)  + \partial_x (\rho^\varepsilon_- w^\varepsilon_- u^\varepsilon_-) = 0 , \label{E2AR_u-} \\
  & & \hspace{-1cm} w^\varepsilon_+= u^\varepsilon_+ + p^\varepsilon(\rho^\varepsilon_+, \rho^\varepsilon_-),   \label{E2AR_w+} \\
  & & \hspace{-1cm} w^\varepsilon_-= -u^\varepsilon_- + p^\varepsilon(\rho^\varepsilon_-, \rho^\varepsilon_+) . \label{E2AR_w-} 
\end{eqnarray*}

\subsubsection{Congestion model with abrupt transitions}
\label{subsubsec_2WAR_density_constraint_incompressibility}

This case corresponds to the formal limit $\varepsilon \to 0$ of the previous model. Suppose that $\rho^\varepsilon \to \rho < \rho^*$. In this case, $Q^\varepsilon(\rho^\varepsilon_\pm,\rho^\varepsilon_\mp) \to 0$ and we recover a 2W-AR model associated to the pressure $P(\rho)$:   

\begin{eqnarray*}
  & & \hspace{-1cm} \partial_t \rho^0_+ + \partial_x (\rho^0_+ u^0_+) = 0 , \label{02AR_n+_NC} \\
  & & \hspace{-1cm} \partial_t \rho^0_- + \partial_x (\rho^0_- u^0_-) = 0 , \label{02AR_n-_NC} \\
  & & \hspace{-1cm} \partial_t (\rho^0_+ w^0_+)  + \partial_x (\rho^0_+ w^0_+ u^0_+) = 0 , \label{02AR_u+_NC} \\
  & & \hspace{-1cm} \partial_t (\rho^0_- w^0_-)  + \partial_x (\rho^0_- w^0_- u^0_-) = 0 , \label{02AR_u-_NC} \\
  & & \hspace{-1cm} w^0_+= u^0_+ + P(\rho^0), \quad u^0_+ \geq 0,  \label{02AR_w+_NC} \\
  & & \hspace{-1cm} w^0_-= -u^0_- + P(\rho^0), \quad u^0_- \leq 0 . \label{02AR_w-_NC} 
\end{eqnarray*}

If on the other hand, $\rho^\varepsilon \to \rho^*$, then $Q^\varepsilon (\rho_+^\varepsilon,\rho_-^\varepsilon) \to \bar Q_+$ and $Q^\varepsilon (\rho_-^\varepsilon,\rho_+^\varepsilon) \to \bar Q_-$. Furthermore, following (\ref{2AR_p_blow4}), $\bar Q_+$ and $\bar Q_-$ are related by: 
\begin{eqnarray}
  & & \hspace{-1cm} q(\rho^0_+) \, \bar Q_+ = q(\rho^0_-) \, \bar Q_- .   \label{2AR_p_blow5}
\end{eqnarray}
Therefore, the total pressure is such that $p^\varepsilon (\rho_+^\varepsilon,\rho_-^\varepsilon)  \to \bar p_+$ and $p^\varepsilon (\rho_-^\varepsilon,\rho_+^\varepsilon)  \to \bar p_-$ with $P(\rho^*) \leq \bar p_\pm $ and $\bar p_+$ and $\bar p_-$ related through (\ref{2AR_p_blow5}) (with $\bar Q_\pm$ replaced by $\bar  p_\pm- P(\rho^*)$).

We stress the fact that ${\bar Q_\pm}$ and consequently ${\bar p_\pm}$ are
\underline{not} local function of $\rho_+^0,\,\rho_-^0$ (only the ratio ${\bar Q_+}/{\bar
  Q_-} = q(\rho_-^0)/q(\rho_+^0)$ is a local function of $\rho_+^0,\,\rho_-^0$). Indeed the
value of ${\bar Q_\pm}$ of two different solutions of the model may be different, even if
the local values of $(\rho_+^0,\,\rho_-^0)$ are the same. Therefore, there is no local
function of $(\rho_+^0,\,\rho_-^0)$ which can match the value of ${\bar Q_\pm}$.

Then, in this case, the model becomes: 
\begin{eqnarray}
  & & \hspace{-1cm} \partial_t \rho^0_+ + \partial_x (\rho^0_+ u^0_+) = 0 , \nonumber \\ 
  & & \hspace{-1cm} \partial_t \rho^0_- + \partial_x (\rho^0_- u^0_-) = 0 , \nonumber \\ 
  & & \hspace{-1cm} \partial_t (\rho^0_+ w^0_+)  + \partial_x (\rho^0_+ w^0_+ u^0_+) = 0 , \nonumber \\ 
  & & \hspace{-1cm} \partial_t (\rho^0_- w^0_-)  + \partial_x (\rho^0_- w^0_- u^0_-) = 0 , \nonumber \\ 
  & & \hspace{-1cm} w^0_+= u^0_+ + \bar p_+ \quad \mbox{ with } \quad P(\rho^*) \leq \bar p_+ ,\nonumber \\ 
  & & \hspace{-1cm} w^0_-= -u^0_- + \bar p_-\quad \mbox{ with } \quad P(\rho^*) \leq \bar p_- , \nonumber \\ 
  & & \hspace{-1cm} \rho^0_+ + \rho^0_- = \rho^*,   \label{02AR_constant} \\
  & & \hspace{-1cm} q(\rho^0_+) \, (\bar p_+ - P(\rho^*)) = q(\rho^0_-) \, (\bar p_- - P(\rho^*)) .   \label{02AR_consistency}
\end{eqnarray}
Relations (\ref{02AR_constant}) and (\ref{02AR_consistency}) furnish the two supplementary
relations which allow us to compute the two additional quantities $\bar
p_+$ and $\bar p_-$. The last relation (\ref{02AR_consistency}) specifies how, at
congestion, the left and right going pedestrians share the available space. We see that
this sharing relation depends upon the choice of the function $q$. Obviously, $q$ is an
input of the model which must be determined from the experimental measurements. If some
flow asymmetry must be taken into account (like if one crowd is more pushy than the other
one), different functions $q_+(\rho_+)$ and $q_-(\rho_-)$ can be used.

This model is a system of first-order differential equations in which the fluxes are
implicitly determined by the constraint (\ref{02AR_constant}). As a consequence of this
constraint, the total particle flux $\rho^0_+ u^0_+ + \rho^0_- u^0_-$ is constant within
the congestion region. We note the difference between this constrained model and the
constrained 1W-AR model (see section
\ref{subsubsec_AR_density_constraint_incompressibility}). In the 1W-AR model, there was a
single unknown congestion pressure $\bar p$ and a single density constraint $\rho =
\rho^*$. In the 2W-AR model, there are two congestion pressures $\bar p_+$ and
$\bar p_-$, which play a similar role in the dynamics of their associated category of
pedestrians. However, there is still a single density constraint, acting on the total
density $\rho_+ + \rho_- = \rho^*$. The additional condition which allows for the
computation of the two congestion pressures is provided by the 'space-sharing' constraint
(\ref{02AR_consistency}). The two constraints express very different physical requirements
and must be combined in order to find the two congestion pressures which, themselves, have
a symmetric role.

\subsubsection{Introduction of the congestion constraint in the 2W-CAR model}
\label{subsubsec_T2AR_density_constraint}

\mbox{}

\noindent {\it (i) Congestion model with smooth transitions.} The smooth pressure relations (\ref{2AR_p_blow})-(\ref{2AR_p_blow3}) can be used for the 2W-CAR model. With this pressure relation, we anticipate that the bound $\rho \leq \rho^*$ is enforced. 

\medskip

\noindent
{\it (ii) Congestion model with abrupt transitions.} If the limit $\varepsilon \to 0$ is considered, then, the limit model in the uncongested region remains of the same form, i.e. is given by (\ref{2ARPed_n+}), (\ref{2ARPed_n-}) with the pressure given by $p(\rho^0_+ , \rho^0_-)= P(\rho^0_+ + \rho^0_-)$. In the congested region, using the same arguments as in section \ref{subsubsec_2WAR_density_constraint_incompressibility}, we find that $(\rho^0_+, \rho^0_-)$ satisfies:  
\begin{eqnarray}
  & & \hspace{-1cm} \partial_t \rho^0_+ + \partial_x (\rho^0_+ (V - \bar p_+) ) = 0 ,
  \nonumber \\ 
  & & \hspace{-1cm} \partial_t \rho^0_- - \partial_x (\rho^0_- (V - \bar p_-) ) = 0 , \nonumber \\ 
  & & \hspace{-1cm} \rho^0_+ + \rho^0_- = \rho^*,   \label{02ARPed_constant} \\
  & & \hspace{-1cm} q(\rho^0_+) \, (\bar p_+ - P(\rho^*)) = q(\rho^0_-) \, (\bar p_- - P(\rho^*)) .   \label{02ARPed_consistency}
\end{eqnarray}
Again, this model gives rise to a system of first order differential equations in which the fluxes are implicitly determined by the constraints (\ref{02ARPed_constant}), (\ref{02ARPed_consistency}). As a consequence of this constraint, the total particle flux $\rho^0_+ u^0_+ + \rho^0_- u^0_-$ (where $u^0_\pm = V - \bar p_\pm$) is constant within the congestion region.

\section{Two-way multi-lane traffic model}
\label{mlane}

\subsection{A Two-way multi-lane Aw-Rascle model of pedestrians}
\label{subsec_mlane_principles}

We now consider a multi-lane model to describe the structure of the flow in the cross sectional direction to the corridor. The models presented so far considered averaged quantities in the cross section of the corridor. However, it is a well observed phenomenon that two-way pedestrian flow presents interesting spontaneous lane structures (see e.g. \cite{Burstedde_2001}), with a preferential side depending on sociological behavior: pedestrians show a preference to the right side in western countries, while the preference is to the left in Japan for instance. In order to allow for a description of the cross-section of the flow, we discretize space in this cross-sectional direction and suppose that pedestrians walk along discrete lanes, like cars on a freeway, with lane changing probabilities depending on the state of the downwind flow. In this way, we design a model which may, if the parameters are suitable chosen, exhibit the spontaneous emergence of a structuration of the flow into lanes. We stress however, that the lanes in our model must be viewed as a mere spatial discretization and that spontaneously emerging pedestrian lanes may actually consist of several contiguous discrete lanes of our model. 

Let $k \in {\mathbb Z}$ be the lane index. So far, we
consider an infinite number of lanes. Of course, there is a maximal number of $K$ lanes
and $k \in \{1, \ldots, K\}$. Extra-conditions due to the finiteness of the number of
lanes are discarded here for simplicity. For each of the lane, we write a 2W-AR model in the
form described in section \ref{1lane}, supplemented by lane-changing source terms.

\begin{definition} {\bf (ML-AR model)} For any index $k \in {\mathbb Z}$, let
  $\rho_{k,\pm}$ the density of pedestrians in the $k$-th lane, $u_{k,\pm}$ their
  velocity, $w_{k,\pm}$ their desired velocity and $p_{k}$ a pressure term, with an
  index $+$ for the right-going pedestrians and $-$ for the left-going ones. The ML-AR
  model is given by:
  \begin{eqnarray}
    & & \hspace{-1cm} \partial_t \rho_{k,+} + \partial_x (\rho_{k,+} \, u_{k,+}) = S_{k,+} , \label{kAR_n+} \\
    & & \hspace{-1cm} \partial_t \rho_{k,-} + \partial_x (\rho_{k,-} \,  u_{k,-}) = S_{k,-} , \label{kAR_n-} \\
    & & \hspace{-1cm} \partial_t (\rho_{k,+} \,  w_{k,+})  + \partial_x (\rho_{k,+} \,  w_{k,+} \,  u_{k,+}) = R_{k,+}   , \label{kAR_u+} \\
    & & \hspace{-1cm} \partial_t (\rho_{k,-}  \, w_{k,-})  + \partial_x (\rho_{k,-} \,  w_{k,-}  \, u_{k,-}) = R_{k,-}   , \label{kAR_u-} \\
    & & \hspace{-1cm} w_{k,+}= u_{k,+} + p_k(\rho_{k,+}, \rho_{k,-}),  \label{kAR_w+} \\
    & & \hspace{-1cm} w_{k,-}= -u_{k,-} + p_k(\rho_{k,-}, \rho_{k,+}) . \label{kAR_w-} 
  \end{eqnarray}
  where $S_{k,\pm}$ and $R_{k,\pm}$ are source terms coming from the lane-changing transition rates. 
\end{definition}

We allow for different pressure relations in the different lanes, to take into account for instance that the behavior of the pedestrians may be more aggressive in the fast lanes than in the slow ones, or to take into account that circulation along the walls may be different than in the middle of the corridor. This point must be assessed by comparisons with the experiments.  We specify the pressure relation in each lane in the form of (\ref{2AR_p_blow}), (\ref{2AR_p_blow3}) with parameter values depending on $k$. 

We denote by 
\begin{eqnarray*}
  & & \hspace{-1cm} \rho_{k} = \rho_{k,+} + \rho_{k,-} , \label{kAR_rho} 
\end{eqnarray*}
the total density on the $k$-th lane. We assume that the congestion density $\rho^*$ is the same for all lanes (this assumption can obviously be relaxed).

\subsection{Interaction terms in the multi-lane model}
\label{subsec_mlane_sources}

We assume that pedestrians prefer to change lane than to reduce their speed, i.e. they change lane if they feel that the offset velocity of their lane (i.e. $p_k(\rho_{k,+}, \rho_{k,-})$ in the case of right-going pedestrians on lane $k$) increases. If facing such an increase, right-going pedestrians change from lane $k$ to lanes $k \pm 1$ (not changing their direction of motion) with rates $\lambda_{k \to k\pm 1}^{+}$. Similarly, these rates are $\lambda_{k \to k\pm 1}^{-}$ for left-going pedestrians. These rates increase with the value of $(d/dt)_{k,+} (p_k(\rho_{k,+}, \rho_{k,-}))$ for $\lambda_{k \to k\pm 1}^{+}$ and with $(d/dt)_{k,-} (p_k(\rho_{k,-}, \rho_{k,+}))$ for $\lambda_{k \to k\pm 1}^{-}$ to indicate that the lane changing probability is increased when an increase of the downstream density is detected. We have denoted by $(d/dt)_{k,\pm}$ the material derivatives for particles moving on the $k$-th lane in the positive or negative direction: $(d/dt)_{k,\pm} = \partial_t + u_{k,\pm} \partial_x$. Strongly congested lanes do not attract new pedestrians. Therefore, $\lambda_{k \to k+ 1}^{+}$ is also a decreasing functions of $\rho_{k+1}$ which vanishes at congestion, when $\rho_{k+1} = \rho^*$. Similarly, $\lambda_{k \to k+ 1}^{-}$ is decreasing with $\rho_{k+1}$ and vanishes at congestion $\rho_{k+1} = \rho^*$ and $\lambda_{k \to k- 1}^{\pm}$ decreases with $\rho_{k-1}$ and vanishes at congestion $\rho_{k-1} = \rho^*$. 

Given these assumptions on the transition rates, the lane-changing source terms for the density equations are written:
\begin{eqnarray}
  & & \hspace{-1.2cm} S_{k,\alpha} =  \lambda_{k+1 \to k}^{\alpha} \, \rho_{k+1,\alpha} + \lambda_{k-1 \to k}^{\alpha} \,\rho_{k-1,\alpha}  - (\lambda_{k \to k + 1}^{\alpha} + \lambda_{k \to k -  1}^{\alpha} ) \rho_{k,\alpha}, \quad \alpha = \pm  \label{S_k_al_1}.
\end{eqnarray}
It is easy to see that this formulation gives:
\begin{eqnarray*}
  & &  \hspace{-1cm} \sum_{k \in {\mathbb Z}}  S_{k,\alpha} = 0, \quad \alpha = \pm
\end{eqnarray*}
which implies the balance equation of the total number of particles moving in a given direction: 
\begin{eqnarray*}
  & &  \hspace{-1cm} \partial_t \rho_\alpha + \partial_x j_\alpha = 0, \quad \rho_\alpha = \sum_{k \in {\mathbb Z}} \rho_{k,\alpha}, \quad j_\alpha = \sum_{k \in {\mathbb Z}} \rho_{k,\alpha} u_{k,\alpha}, \quad \alpha = \pm. 
  \label{cons_rho_mlane}
\end{eqnarray*}

Concerning the rates $R_{k,\pm}$, we consider that $w_{k,\pm}$ being a Lagrangian quantity, the quantities $\rho_{k,\pm} w_{k,\pm}$ vary according to the same rates as the densities themselves. Hence, we let: 
\begin{eqnarray}
  & & \hspace{-1cm} R_{k,\alpha} =  \lambda_{k+1 \to k}^{\alpha} \, \rho_{k+1,\alpha} \, w_{k+1,\alpha}\,+\, \lambda_{k-1 \to k}^{\alpha} \,\rho_{k-1,\alpha} \, w_{k-1,\alpha} \, \nonumber \\
  & & \hspace{3.5cm}   - (\lambda_{k \to k + 1}^{\alpha} + \lambda_{k \to k -  1}^{\alpha} ) \rho_{k,\alpha} w_{k,\alpha}, \quad \alpha = \pm.  \label{R_k_al_1} 
\end{eqnarray}
The material derivatives of $w_{k,\pm}$ satisfy:
\begin{eqnarray*}
  & & \hspace{-1cm} \left( \frac{d w_{k,+}}{dt} \right)_{k,+} \!\!\! : = \partial_t w_{k,+}  + u_{k,+} \, \partial_x w_{k,+} = \frac{1}{\rho_{k,+}} (R_{k,+} - w_{k,+} \,  S_{k,+})  = \nonumber \\
  & & \hspace{-.2cm} = \lambda_{k+1 \to k}^{+} \, \frac{\rho_{k+1,+}}{\rho_{k,+}} (
  w_{k+1,+} \!-\! w_{k,+} ) + \lambda_{k-1 \to k}^{+} \, \frac{\rho_{k-1,+}}{\rho_{k,+}} (  w_{k-1,+} \!-\! w_{k,+} )
  , \label{kAR_w+_lag} \\
  & & \hspace{-1cm} \left( \frac{d w_{k,-}}{dt} \right)_{k,-} \!\!\!  : = \partial_t w_{k,-}  + u_{k,-} \, \partial_x w_{k,-} = \frac{1}{\rho_{k,-}} (R_{k,-} - w_{k,-} \,  S_{k,-})  = \nonumber \\
  & & \hspace{-.2cm} = \lambda_{k+1 \to k}^{-} \, \frac{\rho_{k+1,-}}{\rho_{k,-}} (
  w_{k+1,-} \!-\! w_{k,-} ) + \lambda_{k-1 \to k}^{-} \, \frac{\rho_{k-1,-}}{\rho_{k,-}} (  w_{k-1,-} \!-\! w_{k,-} )
  . \label{kAR_w-_lag} 
\end{eqnarray*}
The right-hand sides of these equations are not zero because the arrival of pedestrians from different lanes with a different preferred velocity modifies the average preferred velocity.

\subsection{The 'constant desired velocity version' of the two-way multi-lane Aw-Rascle model of pedestrians}
\label{subsec_mlane_toy}

To construct the constant desired velocity Aw-Rascle model for two-way multi-lane pedestrian traffic (ML-CAR model), we must set 
\begin{eqnarray}
  & & \hspace{-1cm} w_{k,+} = w_{k,-}= V  , \label{kARPed_w=v} 
\end{eqnarray}
and 
\begin{eqnarray}
  & & \hspace{-1cm} u_{k,+} = V  - p(\rho_{k,+},\rho_{k,-})  , \quad u_{k,-} = - V  + p(\rho_{k,-},\rho_{k,+})  . \label{kARPed_u} 
\end{eqnarray}
We can check in this case that $S_{k,\pm}$ and $R_{k,\pm}$ have been defined in a coherent
way by (\ref{S_k_al_1}) and (\ref{R_k_al_1}), i.e. that they are such that equations
(\ref{kAR_n+}-\ref{kAR_n-}) and (\ref{kAR_u+}-\ref{kAR_u-}) become equivalent. The corresponding model is written:

\begin{definition} {\bf (ML-CAR model)}
  Let $\rho_{k,\pm}$ the density of pedestrians in the $k$-th lane, $V$ the constant
  desired velocity of pedestrian and $p_k$ the pressure term. The ML-CAR model is given
  by:
  \begin{eqnarray*}
    & & \hspace{-1cm} \partial_t \rho_{k,+} + \partial_x \Big(\rho_{k,+} (V - p_k(\rho_{k,+},\rho_{k,-})) \Big) = S_{k,+} , \label{kARPed_n+} \\
    & & \hspace{-1cm} \partial_t \rho_{k,-} - \partial_x \Big(\rho_{k,-} (V - p_k(\rho_{k,-},\rho_{k,+})) \Big) = S_{k,-} , \label{kARPed_n-} 
  \end{eqnarray*}
  where $S_{k,\pm}$ is given by (\ref{S_k_al_1}).  
\end{definition}

The features of this model are those of the two-way, one-lane CAR model of section \ref{subsec_2TAR_model}, combined with the features of the source terms $S_{k,\pm}$ as outlined in section \ref{subsec_mlane_sources}.

\subsection{Introduction of the congestion constraint in the multi-lane ML-AR model}
\label{subsec_kWAR_density_constraint}

\subsubsection{Congestion model with smooth transitions}
\label{subsubsec_kWAR_density_constraint_smooth}

The prescription for the pressure functions $p_k$ are the same as in section \ref{subsubsec_2WAR_density_constraint_smooth}, except for a possible $k$-dependence of the constants, namely: 
\begin{eqnarray*}
  & & \hspace{-1cm} p_k(\rho_{k,+}, \rho_{k,-}) = p_k^\varepsilon (\rho_{k,+},\rho_{k,-}) = P_k(\rho_k) + Q_k^\varepsilon(\rho_{k,+},\rho_{k,-}) ,  \label{kAR_p_blow} \\
  & & \hspace{-1cm} P_k(\rho_k) = M_k \rho_k^{m_k}, \quad m_k \geq 1,  \label{kAR_p_blow2} \\
  & & \hspace{-1cm}  Q_k^\varepsilon(\rho_{k,+},\rho_{k,-}) = \frac{\varepsilon}{q_k(\rho_{k,+}) \left( \frac{1}{\rho_k} - \frac{1}{\rho^*} \right)^{\gamma_k} }  , \quad \gamma_k > 1.  \label{kAR_p_blow3}
\end{eqnarray*}
With this pressure law, the ML-AR model becomes a perturbation problem. This is indicated by equipping all unknowns with an exponent $\varepsilon$. This pressure relation can be used in the constant desired velocity model of section \ref{subsec_mlane_toy} where all particles move with the same speed $V$.

\subsubsection{Congestion model with abrupt transitions}
\label{subsubsec_kWAR_density_constraint_incompressibility}

This case corresponds to the formal limit $\varepsilon \to 0$ of the previous model. Suppose that $\rho_k^\varepsilon \to \rho_k < \rho^*$. In this case, $Q_k^\varepsilon (\rho_{k,+}^\varepsilon, \rho_{k,-}^\varepsilon) \to 0$ and we recover a ML-AR model associated to the pressure $P_k(\rho_k)$:   
\begin{eqnarray*}
  & & \hspace{-1cm} \partial_t \rho^0_{k,+} + \partial_x (\rho^0_{k,+} \, u^0_{k,+}) = S^0_{k,+} , \label{k0AR_n+_NC} \\
  & & \hspace{-1cm} \partial_t \rho^0_{k,-} + \partial_x (\rho^0_{k,-} \,  u^0_{k,-}) = S^0_{k,-} , \label{k0AR_n-_NC} \\
  & & \hspace{-1cm} \partial_t (\rho^0_{k,+} \,  w^0_{k,+})  + \partial_x (\rho^0_{k,+} \,  w^0_{k,+} \,  u^0_{k,+}) = R^0_{k,+}   , \label{k0AR_u+_NC} \\
  & & \hspace{-1cm} \partial_t (\rho^0_{k,-}  \, w^0_{k,-})  + \partial_x (\rho^0_{k,-} \,  w^0_{k,-}  \, u^0_{k,-}) = R^0_{k,-}   , \label{k0AR_u-_NC} \\
  & & \hspace{-1cm} w^0_{k,+}= u^0_{k,+} + P_k(\rho^0_k),  \label{k0AR_w+_NC} \\
  & & \hspace{-1cm} w^0_{k,-}= -u^0_{k,-} + P_k(\rho^0_k). \label{k0AR_w-_NC} 
\end{eqnarray*}

If on the other hand, $\rho_k^\varepsilon \to \rho^*$, the model becomes: 
\begin{eqnarray*}
  & & \hspace{-1cm} \partial_t \rho^0_{k,+} + \partial_x (\rho^0_{k,+} \, u^0_{k,+}) = S^0_{k,+} , \label{k0AR_n+_C} \\
  & & \hspace{-1cm} \partial_t \rho^0_{k,-} + \partial_x (\rho^0_{k,-} \,  u^0_{k,-}) = S^0_{k,-} , \label{k0AR_n-_C} \\
  & & \hspace{-1cm} \partial_t (\rho^0_{k,+} \,  w^0_{k,+})  + \partial_x (\rho^0_{k,+} \,  w^0_{k,+} \,  u^0_{k,+}) = R^0_{k,+}   , \label{k0AR_u+_C} \\
  & & \hspace{-1cm} \partial_t (\rho^0_{k,-}  \, w^0_{k,-})  + \partial_x (\rho^0_{k,-} \,  w^0_{k,-}  \, u^0_{k,-}) = R^0_{k,-}   , \label{k0AR_u-_C} \\
  & & \hspace{-1cm} w^0_{k,+}= u^0_{k,+} + \bar p_{k,+} \quad \mbox{ with } \quad P(\rho^*) \leq \bar p_{k,+} ,\label{k0AR_w+_C} \\
  & & \hspace{-1cm} w^0_{k,-}= - u^0_{k,-} + \bar p_{k,-} \quad \mbox{ with } \quad P(\rho^*) \leq \bar p_{k,-} , \label{k0AR_w-_C} \\
  & & \hspace{-1cm} \rho^0_{k,+} + \rho^0_{k,-} = \rho^*,   \label{0kAR_constant} \\
  & & \hspace{-1cm} q_k(\rho^0_{k,+}) \, (\bar p_{k,+} - P_k(\rho^*)) = q_k(\rho^0_{k,-}) \, (\bar p_{k,-} - P_k(\rho^*)) .   \label{0kAR_consistency}
\end{eqnarray*}
The source terms are unchanged compared to the $\varepsilon >0$ case, and the interpretation of the model is the same as in section \ref{subsubsec_2WAR_density_constraint_incompressibility}. Performing the limit $\varepsilon \to 0$ in the constant desired velocity model of section \ref{subsec_mlane_toy} follows a similar procedure and is left to the reader.


\section{Study of the diffusive two-way, one-lane CAR models}
\label{sec_math}

In this section, we restrict ourselves to the 2W-CAR model presented in section \ref{subsec_2TAR_model} (i.e. without the introduction of the maximal density constraint), and we investigate the stability of a diffusive perturbation of this model. The goal of this section is to show that the addition of a small diffusivity stabilizes the large wave-numbers in the region of state space where hyperbolicity is lacking. The threshold value of the wave-number below which the instability grows can be related to the size of macroscopic structures observed in real crowd flows. 

\subsection{Theoretical analysis}

We consider the following model which is a slight generalization of the 2W-CAR model: 
\begin{eqnarray}
  & & \hspace{-1cm} \partial_t \rho_+ + \partial_x f(\rho_+,\rho_-)  = \delta \, \partial_x^2 \rho_+ , \label{eq:diff_+} \\
  & & \hspace{-1cm} \partial_t \rho_- - \partial_x f(\rho_-,\rho_+) = \delta \, \partial_x^2 \rho_- . \label{eq:diff_-}
\end{eqnarray}
Typically, for the 2W-CAR model,  $f(\rho_+,\rho_-) = \rho_+ (V - p(\rho_+,\rho_-))$ but we do not restrict ourselves to this simple flux prescription. The assumptions on $f$ are that for fixed $\rho_-$, the function $\rho_+ \to f(\rho_+, \rho_-)$ has the bell-shaped curve of figure \ref{Fig_LWR}, which is characteristic of the LWR flux. For fixed $\rho_+$, the function $\rho_- \to f(\rho_+, \rho_-)$ is just assumed decreasing, meaning that the flux of right-going pedestrians is further reduced as the density of left-going pedestrians increases. By symmetry, the diffusivities $\delta$ are assumed to be the same for the two species of particles. Of course, the diffusivities may depend on the densities themselves, in which case they may be different. But we will discard this possibility here. We denote by  
\begin{eqnarray*}
  & & \hspace{-1cm} 
  \tilde c_{++} = \partial_1 f(\rho_+,\rho_-), \quad \quad
  \tilde c_{+-} = \partial_2 f(\rho_+,\rho_-),  \label{eq:speedu_def1} \\
  & & \hspace{-1cm} 
  \tilde c_{-+} = \partial_2 f(\rho_-,\rho_+), \quad \quad
  \tilde c_{--} = \partial_1 f(\rho_-,\rho_+). \label{eq:speedu_def2}
\end{eqnarray*}
These quantities are related to those defined in section \ref{subsec_2AR_model} for the 2W-AR model by 
\begin{equation} \tilde c_{++} = c_{u_+}, \quad \tilde c_{+-} = - \rho_+ c_{+-}, \quad \tilde c_{--} = - c_{u_-}, \quad \tilde c_{-+} = - \rho_- c_{-+}. 
  \label{eq:constants}
\end{equation}
With the assumptions on $f$, we have that $\tilde c_{+-} \leq 0$, $\tilde c_{-+} \leq 0$, while $\tilde c_{++}$ (resp. $\tilde c_{--}$) decreases from positive to negative values when $\rho_+$ (resp. $\rho_-$) increases.

Any state such that $(\rho_+,\rho_-)$ is independent of $x$ is a stationary solution. 
We study the linearized stability of the system about these uniform steady states. Denoting by $(r_+,r_-)$ its unknowns, the linear system is written:
\begin{eqnarray}
  & & \hspace{-1cm} \partial_t r_+ + \tilde c_{++} \partial_x r_+ + \tilde c_{+-} \partial_x r_-  = \delta \, \partial_x^2 r_+ , \label{eq:lin_diff_+} \\
  & & \hspace{-1cm} \partial_t r_- + \tilde c_{-+} \partial_x r_+ + \tilde c_{--} \partial_x r_-  = \delta \, \partial_x^2 r_- . \label{eq:lin_diff_-}
\end{eqnarray}
We look for solutions which are pure Fourier modes of the form $r_\pm = \bar r_\pm \exp i (\xi x - st)$ where $\bar r_\pm$ is the amplitude of the mode, $\xi$ and $s$ are its wave number and frequency. Inserting the Fourier Ansatz into (\ref{eq:lin_diff_+}), (\ref{eq:lin_diff_-}) leads to a homogeneous linear system for $(\bar r_+,\bar r_-)$. This system has non-trivial solutions if and only if the determinant of the linear system cancels. This results in a relation between $s$ and $\xi$ (the dispersion relation). In this analysis, we restrict to $\xi \in {\mathbb R}$ and are looking for the time stability of the model.  We denote by $\lambda = s / \xi$ the phase velocity of the mode. 

A given mode remains bounded in time, and therefore stable, if and only if the imaginary part of $s$ is non-positive. In the converse situation, the mode is unstable. The system is said linearly stable about the uniform state $(\rho_+,\rho_-)$ if and only if all the modes are stable for all $\xi \in {\mathbb R}$. In the converse situation, the system is unstable, and it is then interesting to look at the range of wave numbers $\xi \in {\mathbb R}$ which generate unstable modes. The following result follows easily from simple calculations: 

\begin{proposition} (i) Suppose $(\rho_+,\rho_-)$ are such that the following condition: 
  \begin{equation}
    \Delta := (\tilde c_{++} + \tilde c_{--})^2 - 4 \tilde c_{+-} \tilde c_{-+} \geq 0 , 
    \label{eq:hyp_cond2}
  \end{equation}
  is satisfied, then the uniform steady state with uniform densities $(\rho_+,\rho_-)$ is linearly stable about $(\rho_+,\rho_-)$. For any given $\xi \in{\mathbb R}$, there exist two modes whose phase velocities $\lambda_\pm(\xi)$ are given by 
  \begin{equation}
    \lambda_{\pm} (\xi) = \frac{1}{2} \left[  \tilde c_{++} - \tilde c_{--} - 2 i \delta \xi  \pm \sqrt \Delta \right]. 
    \label{eq:phase_vel}
  \end{equation}

  \medskip
  \noindent (ii) Suppose that $(\rho_+,\rho_-)$ are such that (\ref{eq:hyp_cond2}) is not true. Then, the uniform steady state with uniform densities $(\rho_+,\rho_-)$ is linearly unstable about $(\rho_+,\rho_-)$. Moreover, we have  
  \begin{equation}
    |\xi| \leq \frac{\sqrt{|\Delta|}}{2\delta} \quad \Longleftrightarrow \quad \exists \mbox{ a mode such that }\mbox{Im} \, \, s > 0 \quad \mbox{(unstable mode)}. 
    \label{eq:unstable}
  \end{equation}
  The phase velocity is given by 
  \begin{equation}
    \lambda_{\pm} (\xi) = \frac{1}{2} \left[  \tilde c_{++} - \tilde c_{--} - 2 i \delta \xi  \pm i \sqrt {|\Delta|} \right]. 
    \label{eq:phase_vel_2}
  \end{equation}

  \label{prop:stab_diffus}
\end{proposition}

\medskip
\noindent
We note that if (\ref{eq:constants}) is inserted in (\ref{eq:hyp_cond2}), we recover (\ref{eq:hyp_cond}). Therefore, the addition of diffusion does not change the criterion for stability or instability. However, in the unstable case, all modes are unstable for the diffusion-free model (this would correspond to $\delta = 0$ in (\ref{eq:unstable})). The addition of a non-zero diffusivity stabilizes the modes corresponding to the small scales (large $\xi$). However, the large scale modes (small $\xi$) remain unstable. We also note that, in the stable case, letting the diffusivity go to zero allows us to recover the characteristic speed of the diffusion-free model (\ref{eq:char_vel_2}). 

For unstable modes, (\ref{eq:phase_vel_2}) provides the typical growth rate $\nu_g$: it is equal to the positive imaginary part of $|\xi| \lambda_+$\, , and given by 
$$ \nu_g = \frac{\sqrt{|\Delta|}}{2} |\xi| - \delta \xi^2. $$
It is maximal for 
$$ |\xi| = \frac{\sqrt{|\Delta|}}{4\delta}. $$
Therefore, the typical length scale $L_s$ of the unstable structures is given by the inverse of this wave-number:
$$ L_s = \frac{4 \delta}{\Delta} , $$
because the other modes, having smaller growth rate, will eventually disappear compared to the amplitude of the dominant one. These length scale $L_s$ and time scale $1/\nu_g$ may be related to observations and provide a way to assess the model and calibrate it against empirical data.

\subsection{Numerical simulations}

In this part, we want to investigate numerically the system
(\ref{eq:diff_+}),(\ref{eq:diff_-}) and in particular we are interested in the profile of
the solutions whether the system is in a hyperbolic region or not.

With this aim, we first fix a flux function $f(\rho_+,\rho_-)$ defined as:
\begin{equation}
  \label{eq:flux_f_simu}
  f(\rho_+,\rho_-) = \rho_+\,\frac{g(\rho_++\rho_-)}{\rho_+ + \rho_-},
\end{equation}
where $g$ is a flux depending on the total density $\rho=\rho_+ + \rho_-$. We choose for $g$ a
simple function increasing on $[0,a]$ and decreasing on $[a,1]$:
\begin{displaymath}
  g(x) = \left\{
    \begin{array}{ll}
      \displaystyle x - \frac{x^2}{2a}  & \text{for } 0 \leq x \leq a \\
      \displaystyle  \frac{a}{2} - \frac{a(a-x)^2}{2(1-a)^2} & \text{for } a \leq x \leq 1 \\
      0 & \text{otherwise}
    \end{array}
  \right.
\end{displaymath}
Note that here, in order to keep the simulations simple, we choose a much smoother
expression for $f$ than the one that was proposed in section
\ref{subsec_2WAR_density_constraint} to enforce the density constraint.  As a result, the
density here can become larger than $\rho^*=1$. 

\begin{figure}[ht]
  \centering
  \includegraphics[scale=.4]{figures/function_fg.eps} \quad
  \includegraphics[scale=.4]{figures/Delta_f.eps}
  \caption{Left figure: the flux function $f(\rho_+,\rho_-)$ (\ref{eq:flux_f_simu}) used in our
    simulations. Right figure: the region of non-hyperbolicity (\ref{eq:hyp_cond2}) of the
    model, e.g. $\Delta<0$ in this region.}
  \label{fig:function_fg}
\end{figure}

In the following, we take the maximum of $g$ to be at $.7$, e.g. $a=.7$. The function $f$
is a decreasing function of $\rho_-$ since $g$ satisfies $g'(x)\leq 1$ and $f$ is zero when the
total mass is greater than $1$, e.g. $f(\rho_+,\rho_-)=0$ if $\rho_+ +\rho_-\geq 1$. We
plot the graph of the function $f$ in figure \ref{fig:function_fg} (left). Then, we
numerically compute $\Delta$ to determine the region where the system is non-hyperbolic
(see figure \ref{fig:function_fg}, right). 

To solve numerically the system (\ref{eq:diff_+}),(\ref{eq:diff_-}), we use a central
scheme \cite{kurganov_new_2000}. With this aim, we consider a uniform grid in space
$\{x_i\}_i$ ($\Delta x=x_{i+1}-x_i$) on a fixed interval $[0,L]$ along with a fixed time
step $\Delta t$. We denote by $U_i^n$ the approximation of $(\rho_+,\rho_-)$ on the cell
$[x_{i-1/2},x_{i+1/2}]$ (with $x_{i+1/2} = x_i+\Delta x/2$) at the time $n\Delta t$. The
numerical scheme consists of the following algorithm:
\begin{displaymath}
  \frac{U_i^{n+1}-U_i^n}{\Delta t} + \frac{1}{\Delta x}\left(F_{i+1/2}-F_{i-1/2}\right) = \delta\,
  \frac{U_{i-1}^n-2U_i^n+U_{i+1}^n}{\Delta x^2}.
\end{displaymath}
Here, $F_{i+1/2}$ denotes the numerical flux at $x_{i+1/2}$ defined as:
\begin{displaymath}
  F_{i+1/2} = \frac{F(U_{i+1/2}^L)+F(U_{i+1/2}^R)}{2} - a_{i+1/2} \frac{U_{i+1/2}^R-U_{i+1/2}^L}{2},
\end{displaymath}
where $F$ is the flux of the system $F(\rho_+,\rho_-)=(f(\rho_+,\rho_-),-f(\rho_-,\rho_+))^T$, the vectors
$U_{i+1/2}^L$ and $U_{i+1/2}^R$ are respectively the left and right value of $(\rho_+,\rho_-)$
at $x_{i+1/2}$ computed using a MUSCL scheme \cite{leveque2002fvm} and $a_{i+1/2}$ is the
maximum eigenvalues (\ref{eq:char_vel_2}) of the system at $x_i$ and $x_{i+1}$:
\begin{displaymath}
  a_{i+1/2} = \max(|\lambda_i^{\pm}|,|\lambda_{i+1}^{\pm}|).
\end{displaymath}
As initial condition, we use a uniform stationary state $(\rho_+,\rho_-)$ perturbed by stochastic noise:
\begin{displaymath}
  \rho_+(0,x) = \rho_+ + \sigma\epsilon_+(x) \quad,\quad \rho_-(0,x) = \rho_- + \sigma\epsilon_-(x),
\end{displaymath}
with $\epsilon_+(x)$ and $\epsilon_-(x)$ two independent white noises and $\sigma$ the
standard deviation of the noise. We use periodic boundary condition for our
simulations. The parameters of our simulations are the following: space mesh $\Delta x=1$,
time step $\Delta t=.2$ (CFL$=.406$), diffusion coefficient $\delta=.4$ and standard
deviation of the noise $\sigma=10^{-2}$. We use periodic boundary conditions.

To illustrate our numerical scheme, we use three different initial conditions. First, we
pick two values for $(\rho_+,\rho_-)$ in the hyperbolic region:
\begin{displaymath}
  \rho_+ = .35 \quad,\quad \rho_-=.3.
\end{displaymath}
The initial datum is plotted on figure \ref{fig:equil_03503} (left). As we can see on figure \ref{fig:equil_03503} (right), the solution stabilizes around the
stationary state $(.35,.3)$.

For our second simulation, we take $(\rho_+,\rho_-)$ in a non-hyperbolic region:
\begin{displaymath}
  \rho_+ = .5 \quad,\quad \rho_-=.3.
\end{displaymath}
The solution does no longer stabilize around the stationary state $(.5,.3)$. On
figure \ref{fig:equil_0503} (left), we observe the apparition of clusters of high density. Each cluster
for $\rho_+$ faces a cluster for $\rho_-$. Moreover, in each cluster, the total mass
$\rho_++\rho_-$ is greater or equal to $1$. Therefore the flux in this region is
zero.
However, due to the diffusion, the solution is not in a stationary state. There is
exchange of mass between the clusters. If we run the solution for a long time, only one
cluster remains (see figure \ref{fig:equil_0503} (right)). In this cluster, we observe that the
profile of $\rho_+$ is concave-down whereas the profile of $\rho_-$ is
concave-up. Consequently, the diffusivity makes $\rho_+$ moving backward and $\rho_-$
moving forward. As a result, all the clusters are moving to the left. However, the concavity
of the solution is puzzling. Numerically, it appears that the concavity of $\rho_+$ and
$\rho_-$ depends on the total mass: the density with higher mass is concave-down and the
density with lower mass is concave-up. But this property has to be understood
analytically.

For the third simulation, we take an initial datum $(\rho_+,\rho_-)$ close to the non-hyperbolic
region:
\begin{displaymath}
  \rho_+ = .4 \quad,\quad \rho_-=.3.
\end{displaymath}
Indeed, we can see on figure \ref{fig:function_fg} (right) that the point $(0.4,0.3)$ almost lies at the border of the non-hyperbolic region. 
The oscillations amplify and clusters of high densities emerge (figure
\ref{fig:equil_0403}, left). However, if we increase the  diffusion coefficient, taking
$\delta=2$ instead of $\delta=.4$, then the solution stabilizes around the stationary
state $(.4,.3)$ as we observe on figure \ref{fig:equil_0403}, right. Therefore,
a large enough diffusion prevents cluster formation.

\begin{figure}[p]
  \centering
  \includegraphics[scale=.3]{figures/simulations/equil_init_03503.eps} \hspace{1cm}
  \includegraphics[scale=.3]{figures/simulations/equil_03503.eps}
  \caption{The initial condition (left figure) and the solution at $t=500$ unit times. The solution
    stabilizes around the stationary state $(.35,.3)$.}
  \label{fig:equil_03503}
\end{figure}

\begin{figure}[p]
  \centering
  \includegraphics[scale=.3]{figures/simulations/equil_0503.eps} \hspace{1cm}
  \includegraphics[scale=.3]{figures/simulations/equil_0503_t105.eps}
  \caption{Starting from the initial state $(.5,.3)$, the initial oscillations amplify to
    create clusters (left figure). After a longer time ($t=10^4$ unit times), only one
    cluster remains (right figure).}
  \label{fig:equil_0503}
\end{figure}

\begin{figure}[p]
  \centering
  \includegraphics[scale=.3]{figures/simulations/equil_0403_diff04.eps} \hspace{1cm}
  \includegraphics[scale=.3]{figures/simulations/equil_0403_diff2.eps}
  \caption{Starting from the initial state $(.4,.3)$, clusters appears once again (left
    figure). However, if we increase the diffusion coefficient ($\delta=2$ instead of $\delta=.4$),
    the solution is stabilizing (right figure).}
  \label{fig:equil_0403}
\end{figure}

The simulations are provided for ilfustration purpose only, and explore what kind of structures the lack of hyperbolicity of the model leads to.  The experimental evidence of the appearance of clusters is difficult to provide since they must occur (if they occur) at very high densities. Experiments in such high density conditions are not possible for obvious safety reasons. The observation of real crowds shows that pedestrians can still move even at very high densities thanks to the spontaneous organization of the flow into lanes. The simple one-dimensional model that is simulated here cannot account for this feature. However, we conjecture that cluster formation can be impeded in the multi-lane model through the introduction of adequate lane-changing probabilities.

\setcounter{equation}{0}
\section{Conclusion}
\label{sec_conclu}

In this work, we have presented extensions of the Aw-Rascle macroscopic model of traffic
flow to two-way multi-lane pedestrian traffic, with a particular emphasis on the study of
the hyperbolicity of the model and the treatment of congestions. 

A first important contribution of the present work is that
two-way models may lose their hyperbolicity in certain conditions and that this may be
linked to the generation of large scale structures in crowd flows. Adding diffusion helps
stabilize the small scale structures and favors the development of large scale structures
which may be related to observations. We have shown numerical simulations which support
this interpretation. 

A second contribution of this work is to provide a methodology to handle the congestion constraint in pedestrian traffic models. Congestion effects reflect the fact that the
density cannot exceed a limit density corresponding to contact between pedestrians.
We have proposed to 
treat them
by a modification of the pressure relation which reduces the pedestrian
velocities when the density reaches this maximal density. If this modification occurs on a
very small range of densities, then, the model exhibits abrupt transitions between
compressible flow (in the uncongested region) and incompressible flow (in the congested
region). 

Mathematically rigorous proofs that these models respect the upper-bound on the total density are
left to future works. Their numerical resolution will require the development specific
techniques such as Asymptotic-Preserving methodologies in order to treat the occurrence of
congestions. Data learning techniques will then be applied to fit the parameters of the
model to experimental data.
Other possible extensions of this work are the development of more complex models such as two-dimensional models, kinetic models allowing for a statistical distribution of velocities or crowd turbulence models with weak compressibility near congestion. Finally, the derivation of approximate equations describing the geometric evolution of the transition interface between the uncongested and congested regions would help understanding the dynamics of these interfaces. 



\medskip

\medskip


\begin{thebibliography}{99}


\bibitem{Al-nasur_2006} S. Al-nasur \& P. Kachroo, {\em A Microscopic-to-Macroscopic Crowd Dynamic model}, 
  Proceedings of the IEEE ITSC 2006, 2006 IEEE Intelligent Transportation Systems Conference
  Toronto, Canada, September 17-20 (2006). 

\bibitem{AKMR} A. Aw, A. Klar, A. Materne, M. Rascle,  {\em Derivation of continuum traffic flow models from microscopic follow-the-leader models}, SIAM J. Appl. Math., {\bf 63} (2002), 259--278. MR1952895 (2003m:35148)

\bibitem{AR} A. Aw, M. Rascle, {\em Resurrection of second order models of traffic flow}, SIAM J. Appl. Math., {\bf 60} (2000), 916--938. MR1750085 (2001a:35111) 

\bibitem{Bellomo_d08} N. Bellomo, C. Dogbe, {\em On the modelling crowd dynamics: from scaling to second order hyperbolic macroscopic models}, Math. Models Methods Appl. Sci.,  {\bf 18} (2008), 1317--1345. MR2438218 (2009g:92074)

\bibitem{benzoni} S. Benzoni-Gavage, R. M. Colombo, {\em An $n$-populations model for traffic flow}, European J. Appl. Math.,  {\bf 14} (2003), 587--612. MR2020123 (2004k:90018)

\bibitem{BDDR} F. Berthelin, P. Degond, M. Delitala, M. Rascle, {\em A model for the formation and evolution of traffic jams}, Arch. Rat. Mech. Anal., {\bf 187} (2008), 185--220. MR2366138 (2008h:90022)

\bibitem{BDLMRR} F. Berthelin, P. Degond, V. Le Blanc, S. Moutari, J. Royer, M. Rascle, {\em A Traffic-Flow Model with Constraints for the Modeling of Traffic Jams}, Math. Models Methods Appl. Sci., {\bf 18, Suppl.} (2008), 1269-1298. MR2438216 (2010f:35233)

\bibitem{Bou_Bre_Cor_Rip}  F. Bouchut, Y. Brenier, J. Cortes, J. F. Ripoll, {\em A hierachy of models for two-phase flows}, J. Nonlinear Sci., {\bf 10} (2000), 639--660. MR1799394 (2001j:76109)

\bibitem{Burstedde_2001} C. Burstedde, K. Klauck, A. Schadschneider, J. Zittarz, {\em Simulation of pedestrian dynamics using a 2-dimensional cellular automaton}, Physica A, {\bf 295} (2001), 507--525. arXiv:cond-mat/0102397

\bibitem{Chalons_2007} C. Chalons, {\em Numerical approximation of a macroscopic model of pedestrian flows}, SIAM J. Sci. Comput., {\bf 29} (2007), 539--555. MR2306257 (2008a:35188)

\bibitem{Colombo_2005} R. M. Colombo, M. D. Rosini, {\em Pedestrian flows and nonclassical shocks},  Math. Methods Appl. Sci., {\bf 28} (2005), 1553--1567. MR2158218 (2006b:90009) 

\bibitem{Dag} C. Daganzo, {\em Requiem for second order fluid approximations of traffic flow}, Transp. Res. B, {\bf 29} (1995), 277--286.

\bibitem{Deg_Del} P. Degond, M. Delitala, {\em Modelling and simulation of vehicular traffic jam formation}, Kinet. Relat. Models, {\bf 1} (2008), 279--293. MR2393278 (2009a:90022)

\bibitem{DHN} P. Degond, J. Hua, L. Navoret, {\em Numerical simulations of the Euler system with congestion constraint}, preprint. arXiv:1008.4045

\bibitem{DT} P. Degond, M. Tang, {\em All speed scheme for the low Mach number limit of the Isentropic Euler equations}, Commun. Comput. Phys., {\bf 10} (2011), 1-31. arXiv:0908.1929

\bibitem{Guo_2008} R. Y. Guo, H. J. Huang, {\em A mobile lattice gas model for simulating pedestrian evacuation}, Physica A, {\bf 387} (2008), 580--586.

\bibitem{Guy09} S. J. Guy, J. Chhugani, C. Kim, N. Satish, M. C. Lin, D. Manocha,
P. Dubey, {\em Clearpath: Highly parallel collision avoidance for multi-agent simulation}, in ACM SIGGRAPH/Eurographics Symposium on Computer Animation (SCA), pp. 177--187 (2009).

\bibitem{Helbing_1991} D. Helbing, {\em A mathematical model for the behavior of pedestrians},  Behavioral Science, {\bf 36} (1991), 298--310.

\bibitem{Helbing_1992} D. Helbing, {\em A fluid dynamic model for the movement of pedestrians}, Complex Systems, {\bf 6} (1992), 391--415. 

\bibitem{Helbing_Molnar_1995} D. Helbing, P. Moln\`ar, {\em Social force model for pedestrian dynamics},  Physical Review E, {\bf 51} (1995), 4282--4286.

\bibitem{Helbing_Molnar_1997} D. Helbing, P. Moln\`ar, {\em Self-organization phenomena in pedestrian crowds} in: F. Schweitzer (ed.) Self-Organization of Complex Structures: From Individual to Collective Dynamics, pp. 569--577, Gordon and Breach, London (1997). 

\bibitem{Henderson_1974} L. F. Henderson, {\em On the fluid mechanics of human crowd motion}, Transportation Research, {\bf 8} (1974), 509--515. 

\bibitem{Hoogendoorn_2003} S. Hoogendoorn, P. H. L. Bovy, {\em Simulation of pedestrian flows by optimal control and differential games}, Optimal Control Appl. Methods, {\bf 24} (2003), 153--172. MR1988582 (2004d:93088)

\bibitem{Hughes_2002} R. L. Hughes, {\em A continuum theory for the flow of pedestrians}, Transportation Research B, {\bf 36} (2002), 507--535. 

\bibitem{Hughes_2003} R. L. Hughes, {\em The flow of human crowds}, Ann. Rev. Fluid Mech., {\bf 35} (2003), 169--182.  

\bibitem{kurganov_new_2000}  A. Kurganov, E. Tadmor, {\em New high-resolution central schemes for nonlinear conservation laws and convection-diffusion equations}, J. Comput. Phys., {\bf 160} (2000), 240--282. MR1763829 (2001c:65102)

\bibitem{leveque2002fvm} R. J. LeVeque, "Finite volume methods for hyperbolic problems", Cambridge University Press (2002).
  
\bibitem{LW} M. J. Lighthill, J. B. Whitham, {\em On kinematic waves. I: flow movement in long rivers. II: A theory of traffic flow on long crowded roads}, Proc. Roy. Soc., {\bf A229} (1955), 281--345. MR0072605 (17,309e) and MR0072606 (17,310a). 

\bibitem{Maury_Roudneff_2010} B. Maury, A. Roudneff-Chupin, F. Santambrogio, {\em A macroscopic crowd motion model of gradient flow type}, Math. Models Methods Appl. Sci., {\bf 20} (2010), 1787--1821. MR2735914

\bibitem{Maury_Venel_2008} B. Maury, J. Venel, {\em A mathematical framework for a crowd motion model}, C. R. Acad. Sci. Paris, Ser. I, {\bf 346} (2008), 1245--1250. MR2473301 (2009m:91150)

\bibitem{Nishinari_2004} K. Nishinari, A. Kirchner, A. Namazi, A. Schadschneider, {\em Extended floor field CA model for evacuation dynamics}, IEICE Transp. Inf. \& Syst., {\bf E87-D} (2004), 726--732. arXiv:cond-mat/0306262

\bibitem{Ondrej10} J. Ond{\v{r}}ej, J. Pettr\'e, A-H. Olivier, S. Donikian, {\em A synthetic-vision based steering approach for crowd simulation}, in SIGGRAPH '10 (2010).

\bibitem{Paris_p_d07} S. Paris, J. Pettr\'e, S. Donikian, {\em Pedestrian reactive navigation for crowd simulation: a predictive approach}, Eurographics, {\bf 26} (2007), 665--674.

\bibitem{Ped1} Pedigree team, {\em Pedestrian flow measurements and analysis in an annular setup}, in preparation. 

\bibitem{Pettre09} J. Pettr\'e, J. Ond{\v{r}}ej, A-H. Olivier, A. Cretual, S. Donikian, {\em Experiment-based modeling, simulation and validation of interactions between virtual walkers}, in SCA '09: Proceedings of the 2009 ACM SIGGRAPH/Eurographics Symposium on Computer Animation, pp.189-198 (2009).

\bibitem{Piccoli_2009} B. Piccoli, A. Tosin, {\em Pedestrian flows in bounded domains with obstacles}, Contin. Mech. Thermodyn., {\bf 21} (2009), 85-107. arXiv:0812.4390

\bibitem{Piccoli_2010} B. Piccoli, A. Tosin, {\em Time-evolving measures and macroscopic modeling of pedestrian flow}, Arch. Ration. Mech. Anal., {\bf 199} (2010), 707--738. arXiv:0811.3383
  
\bibitem{Reynolds99} C. W. Reynolds, {\em Steering behaviors for autonomous characters}, in Proceedings of Game Developers Conference 1999, San Jose, California, pp. 763-782 (1999).

\bibitem{Shvetsov_1999} V. Shvetsov, D. Helbing, {\em Macroscopic dynamics of multi-lane traffic}, Phys. Rev. E {\bf 59} (1999), 6328-6339. arXiv:cond-mat/9906430

\bibitem{Vandenberg_o08} J. van~den Berg, H. Overmars, {\em Planning time-minimal safe paths amidst unpredictably moving obstacles}, Int. Journal on Robotics Research {\bf 27} (2008), 1274--1294.

\bibitem{Weng_2007} W. G. Weng, S. F. Shena, H. Y. Yuana, W. C. Fana, {\em A behavior-based model for pedestrian counter flow}, Physica A {\bf 375} (2007), 668--678.

\bibitem{Zhang} M. Zhang, {\em A non-equilibrium traffic model devoid of gas-like behavior}, Transportation Res. B {\bf 36} (2002), 275--290. 

\end{thebibliography}
\end{document}